\documentclass[twocolumn,aps,prb,psfig,longbibliography,superscriptaddress]{revtex4-2}
\usepackage{graphicx,amssymb,soul,color,amsmath,bm}
\usepackage{epstopdf,hyperref}
\usepackage{orcidlink}
\usepackage{braket,dsfont,nicefrac}
\usepackage[mathcal]{euscript}
\usepackage[normalem]{ulem}
\usepackage[utf8]{inputenc}
\usepackage[normalem]{ulem}
\hypersetup{colorlinks=true,citecolor=blue,linkcolor=magenta}

\sethlcolor{yellow}
\allowdisplaybreaks

\begin{document}

\title{
Spin-charge separation in two-leg $t-J$ ladders
}

\author{Luhang Yang\orcidlink{0000-0002-3591-0595}}
\email{luhangyang@utk.edu}
\affiliation{Department of Physics and Astronomy, University of Tennessee, Knoxville, Tennessee 37996, USA}

\author{Elbio Dagotto\orcidlink{0000-0001-6007-5694}}
\affiliation{Department of Physics and Astronomy, University of Tennessee, Knoxville, Tennessee 37996, USA}
\affiliation{Materials Science and Technology Division, Oak Ridge National Laboratory, Oak Ridge, Tennessee 37831, USA}

\begin{abstract}
Spin–charge separation is a hallmark of one-dimensional fermionic systems, yet its realization in higher dimensions remains an open question. To address this issue, we investigate a two-leg $t-J$ ladder using the density matrix renormalization group (DMRG) method and its time-dependent extension. By analyzing ground-state correlations and single-particle removal spectra, we systematically examine the effects of plaquette diagonal hopping, spin exchange, and hole doping. Within appropriate parameter regimes, these factors drive the system from the well-known Luther Emery  phase, with gapped spin and gapless charge modes, into a Luttinger liquid phase characterized by gapless spin and charge excitations, where signatures of spin–charge separation emerge. In combination with previous studies using exact diagonalization, our results provide evidence that spin–charge separation may persist in wider ladder systems.

\end{abstract}

\maketitle

\section{Introduction}

Spin–charge separation (SCS) is a typical example of non-Fermi-liquid behavior in interacting electron systems \cite{kivelson1990,murdy1994,zhao1995,si1997,si1998,Nagaosa1998,balents2000}. It was originally proposed and experimentally observed in one-dimensional materials \cite{kim1996observation,sing2003,auslaender2005spin,kim2006distinct,jompol2009probing} and has since been explored in certain quasi-one-dimensional and higher-dimensional systems \cite{Elbio_ladder,elbio_2D, elbio1999,elbio2000,w23h-dhrk,hyun2025}. Recently, SCS has attracted increasing interest for its potential applications in spintronics \cite{linder2015superconducting} and electronic devices \cite{jompol2009probing,rossi2025graphene}, as well as cold-atom studies \cite{senaratne2022spin}. 

In one-dimensional electron chains, SCS is well understood within the framework of Tomonaga–Luttinger liquid (TLL) theory, which uses bosonization and a linearization of the dispersion near the Fermi points \cite{giamarchi2003quantum}. Within this description, spin and charge degrees of freedom emerge as deconfined excitations.
In contrast, the two-dimensional square lattice in Hubbard-like models exhibits long-range magnetic order near half filling, which strongly confines spin and charge degrees of freedom, thus SCS is widely considered as absent in this case. As an intermediate setting between one and two dimensions, ladder systems provide a useful platform for understanding the relationships among different instabilities while remaining computationally feasible \cite{elbio1992,elbio1996,elbio_1999_review}. Indications of SCS have been reported in ladders under appropriate conditions, such as the inclusion of additional hopping terms \cite{Elbio_ladder,elbio_2D}. More recently, frustrated lattices, including honeycomb and kagome geometries have also been suggested to host SCS-like behavior \cite{lauchili2004,miao2025}. These observations lie beyond the conventional Tomonaga–Luttinger liquid (TLL) framework and call for a broader understanding of the conditions under which SCS can emerge.

While the quasi-long range order of antiferromagnetism is not the direct origin of spin–charge separation within the TLL framework, it plays a crucial role in determining the nature of spin excitations. In one-dimensional systems, quasi-long-range antiferromagnetic (AFM) correlations give rise to gapless spinon excitations. At the same time, these correlations are not strong enough to confine spin and charge, allowing for their separation.

In contrast, two-leg ladder systems exhibit qualitatively different behavior. At half filling, multiple studies \cite{Elbio_ladder,elbio_2D, elbio1999} have shown that the ground state can be qualitatively described as a rung-singlet state, with spin singlets at every rung. This state has a finite spin gap. In the limit where the superexchange $J_{rung}$ is much larger than the leg superexchange $J_{leg}$, the dominance of the rung-singlet state becomes exact.
As a result, separating spin and charge degrees of freedom requires breaking the spin bound state on the rung (i.e. breaking the singlet), which costs a finite energy. Consequently, it is generally believed that SCS is absent in two-leg ladders at half filling for the Hubbard or $t-J$ models with nearest-neighbors (NN) hopping.
By introducing next-nearest-neighbor (NNN) hopping and/or hole doping, the ground state of the two-leg ladder can be significantly modified. In particular, for {\it negative} NNN hopping, relative to the NN value, we show that the spin sector becomes gapless both for the undoped and low-hole doping cases, and the system enters a Luttinger liquid phase \cite{my_3band}, which opens the possibility for spin–charge separation (SCS). 

To identify the key ingredients that favor the emergence of SCS, we use numerical methods to investigate the nature of single-particle excitations in a two-leg 
$t-J$ ladder with NNN hopping. We aim to provide insights into the mechanisms underlying the emergence of SCS and its possible realization in more general lattice systems.

\section{Model and Method}

We study the $t_1-t_2-J$ model on two leg ladders, of which the Hamiltonian is:
\begin{eqnarray}
\nonumber & H_{t_1-t_2-J} = -t_1\sum_{\mathbf{r},\sigma}(c_{\mathbf{r},\sigma}^\dagger c_{\mathbf{r}+\hat{\delta_1},\sigma} + h.c.) \\ 
\nonumber & - t_2\sum_{\mathbf{r},\sigma}(c_{\mathbf{r},\sigma}^\dagger c_{\mathbf{r}+\hat{\delta_2},\sigma} + h.c.) \\ 
& +  J\sum_{\mathbf{r}}(\vec{S}_\mathbf{r}\cdot \vec{S}_{\mathbf{r}+\hat{\delta_1}} 
    - \cfrac{1}{4} n_\mathbf{r}n_{\mathbf{r}+\hat{\delta_1}})
\label{tJ}
\end{eqnarray}
where $\vec{S}_\mathbf{r}$ is the spin $S=1/2$ operator at position $\mathbf{r}$, and $J$ parametrizes the spin exchange, $t_1$ represents NN hopping, and $t_2$ is the parameter for NNN hopping. We set $t_1=1$ throughout our study, thus the use of $t_2/t_1$ and $t_2$ are interchangeable. The distances are given by the lattice vectors $\hat{\delta_1} = (0,1)$ or $(1,0)$ for NN hopping, and $\hat{\delta_2} = (1,1)$ or $(1,-1)$ for NNN hopping.
We adopt small values of $J$ ($=0.2$ and $0.4$) so that this $t-J$ model can represent the physics of the hole-doped Hubbard model with large $U$ ($\simeq 4t_1^2/J$) \cite{mapping_hubbard_tJ}.

\begin{figure}
	\centering
  \includegraphics[width=0.4\textwidth]{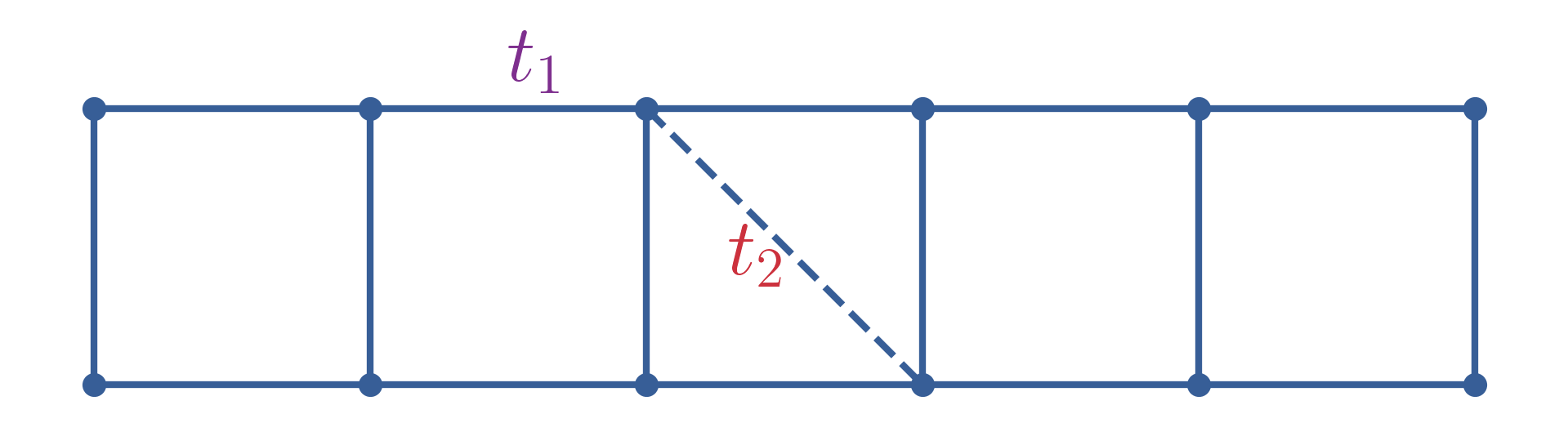} 	
  \caption{Lattice structure of the two-leg ladder.} \label{fig:lattice}
\end{figure}

We use the density matrix renormalization group (DMRG) method \cite{White1992,white1993} to investigate the ground-state properties of two-leg ladders with lengths ranging from 16 to 64 sites along the leg ($x$) direction. The bond dimension is taken up to $m=1200$, ensuring a truncation error no larger than $10^{-6}$.
To probe dynamical properties, we use time-dependent DMRG (tDMRG) \cite{white2004,daley2004,vietri,Paeckel2019} to compute the single-particle removal spectral function of the $t_1-t_2-J$ ladder. The tDMRG calculations are implemented by grouping each rung into a composite site and applying a Suzuki–Trotter decomposition for the time-evolution operator.

We compute the real-time correlation functions $\langle c^\dagger_i(t)c_j(0)\rangle$ with a time step $\tau=0.05$ up to a maximum time $T=60$. The momentum and frequency resolved spectra are then obtained via Fourier transformation, supplemented by a Lorentzian broadening of width $\epsilon=0.1$ to reduce finite-time window artifacts, such as ringing~\cite{white2004}.

\section{Effects of $t_2$ and $J$}

The half-filled two-leg ladder with only nearest-neighbor hopping and strong rung coupling has a ground state dominated  by rung 
singlets \cite{elbio1992,elbio1996,NOACK1996281}. By introducing a single hole, and if the rung superexchange is robust, then the hole and the spin on the same rung form a bound state (that we refer to as ``polaron'' in this publication) that propagates as a composite object. Because the rung-singlet state is believed to provide a qualitatively good description for all coupling (unless the chains fully decouple from each other), then this perspective expresses that the polarons merely change their size as the superexchange decreases, but the hole and spin cloud remain together. In this regime, the system behaves similarly to a band insulator, exhibiting a cosine-like dispersion in the photoemission spectrum \cite{Adrian_hubbard_ladder}.
As the rung coupling decreases toward the isotropic limit (namely when the $J$ of rungs and legs are the same, and the NN hoppings of rungs and legs are the same), the system can undergo a crossover from a band insulator (strong-rung limit) to a spin-liquid regime (strong-leg limit), depending on parameters such as doping or plaquette diagonal hopping (also often referred to as {\it t'}).

In this process, the spin correlation length along the leg direction increases, and the polaron extends over multiple rungs and legs. Consequently, hole motion can be characterized by two distinct modes: in one, the holon remains bounded with the spin excitations, giving rise to a coherent peak in the spectral function; in the other, the holon propagates through a background of incoherent spin fluctuations in the near range, behaving effectively as a spinless fermion in a one-dimensional chain \cite{Adrian_hubbard_ladder}.
The inclusion of a negative NNN hopping is believed to enhance the mobility of the holon along the leg direction. An earlier Lanczos study reported signatures of spin-charge separation at least in the short range, and demonstrated that the holon motion under these conditions can be approximated by that of an effective one-dimensional chain \cite{Elbio_ladder}.

\begin{figure}
	\centering
\includegraphics[width=0.4\textwidth]{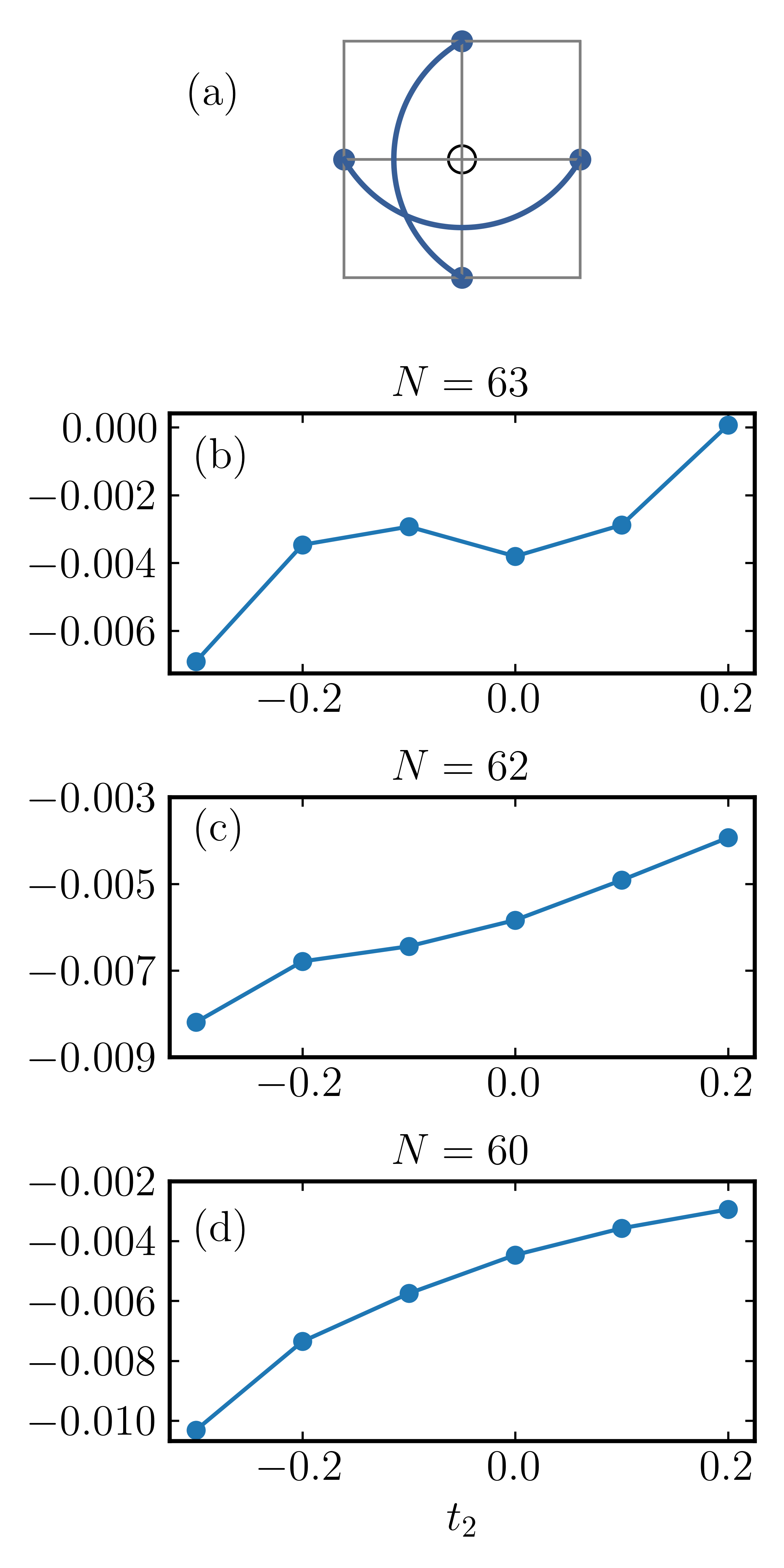}
  \caption{(a): Sketch of spin-spin correlations from Ref.~\cite{Elbio_ladder}, illustrating the concept of spin-charge separation via
the formation of spin singlets across the hole. (b)-(c): Spin-spin correlations across a hole as a function of $t_2$, for different doping densities. The spins considered here are located on the sites adjacent to the hole along the same leg, and the correlation is defined in the vertical axis label. These results are obtained using DMRG for ladders of length $Lx=32$ with $J=0.2$.} \label{fig:szsz_hole}
\end{figure}

To elucidate how the NNN hopping affects the low-energy dynamics, we first compute the ground state spin–spin correlations across a hole. 
This quantity was introduced years ago as a way to judge the degree of spin-charge separation \cite{Elbio_ladder}. 
Figure~\ref{fig:szsz_hole} (a), reproduced from Ref.~\cite{Elbio_ladder}, illustrates the qualitative concept introduced in that early work, providing a local framework in which holes can become decoupled from spins. The figure depicts a hole at the center, with the surrounding spins forming spin singlets. When the spins surrounding a hole are coupled in spin singlets, then the movement of the hole decouples 
from the spins: as the hole moves, the spins adapt, keeping the singlet formation across the hole. 

While in ladders or two-dimensional geometries, this is a local approximate concept that requires computational testing for quantitative confirmation, in the limit of
infinite Hubbard $U$ using the one-dimensional Hubbard model, the idea becomes exact. In this limit, 
holes and spins are fully decoupled and the many-body wave function can be written exactly as the product 
of the spin sector wave function, as if the holes were not there, times the
wave function of the holons, as if the spins were not there \cite{ogata-shiba}. Because the spin sector in this limit is
AFM at least at short distances, then the spins couple antiferromagnetically across each hole.

Previous computational tests using small clusters \cite{Elbio_ladder} suggest that this concept
remains approximately valid in ladders but only after the addition of nearest-neighbor hopping with the proper sign.
Namely the two spins next to the hole are antiferromagnetically 
coupled to one another depending on values of parameters in the Hamiltonian. This allows the hole to move without
fighting against the spin background \cite{Elbio_ladder,elbio_2D}, 
as it occurs in the Ogata-Shiba limit of the Hubbard chain \cite{ogata-shiba}.
In the present study, we investigate whether the across-the-hole antiferromagnetism concept remains valid for two-leg ladders within a specific 
parameter regime when employing both DMRG and tDMRG techniques, improving the cluster sizes accesible to exact diagonalization and adding the dynamical component. 
Specifically, we study
dynamical spectral functions that support the tendency to spin charge separation 
in special regions of parameter space in the extended $t-J$ model, as discussed later in this publication.

To address the spin across the hole concept, we focus on the observable
$\langle S^z(L_x/2-1,0)n_h(L_x/2,0)S^z(L_x/2+1,0)\rangle$, as a function of $t_2$. 
Via the number operator $n_h$, we make sure that the hole is located at the center of the ladder, denoted by $(L_x/2,0)$, to minimize
edge effects. The results for different doping densities are shown in Fig.~\ref{fig:szsz_hole} (b)-(d). The correlation is always negative, indicating that the spins around the hole are antiferromagnetically aligned, but with different strengths. More specifically, as $t_2$ increases toward positive values, the AFM coupling magnitude decreases. As $t_2$ increases in magnitude towards more negative values, the AFM coupling increases. 
If the spinon and holon are strongly confined, the spins around the hole were found to be ferromagnetically aligned \cite{my_tJ_chain}. In contrast, if they deconfine, the spin-spin correlations around the hole should become antiferromagnetic. Therefore, the observed enhancement of the AFM correlations accross the hole as the magnitude of negative $t_2$ increases is qualitatively related to spin-charge separation, as conjectured in previous 
publications using small clusters \cite{Elbio_ladder}.

\begin{figure}
	\centering
  \includegraphics[width=0.5\textwidth]{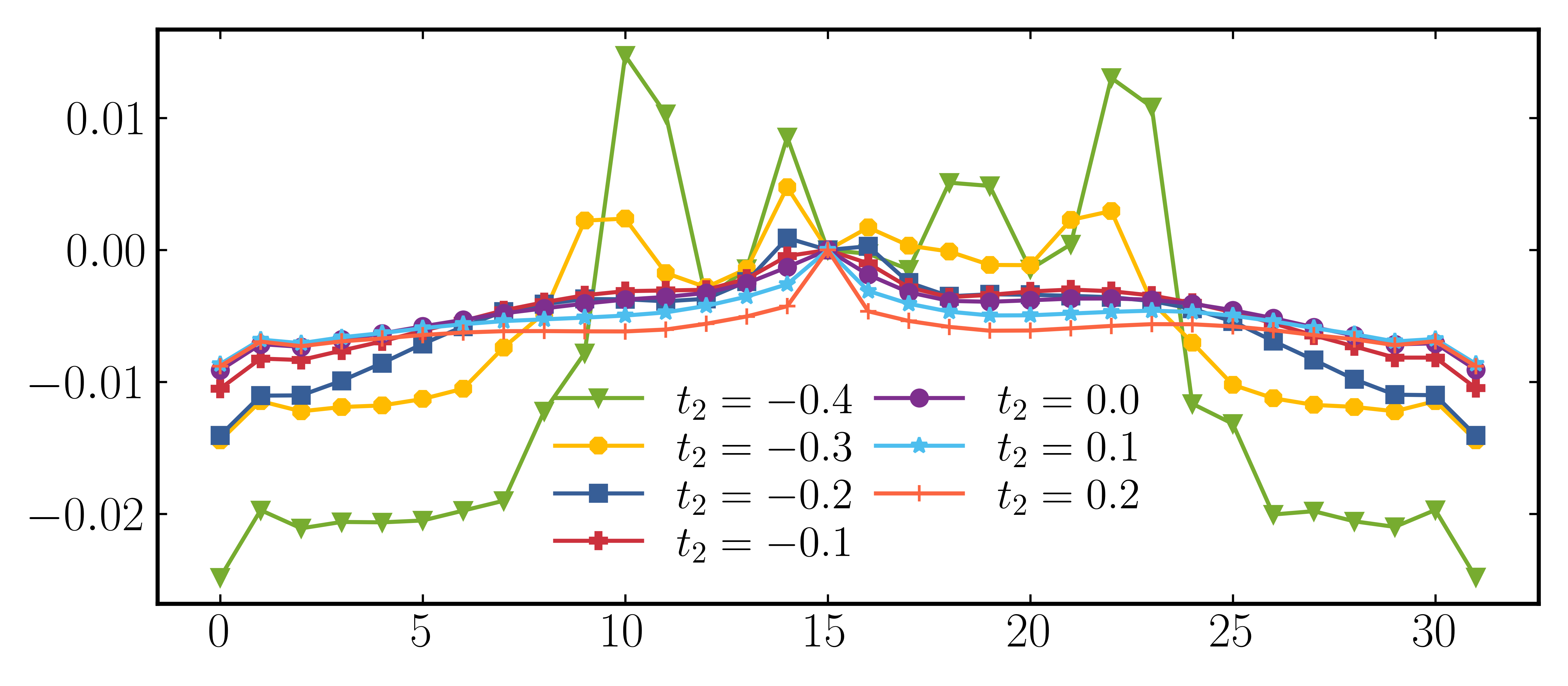} 
  \includegraphics[width=0.5\textwidth]{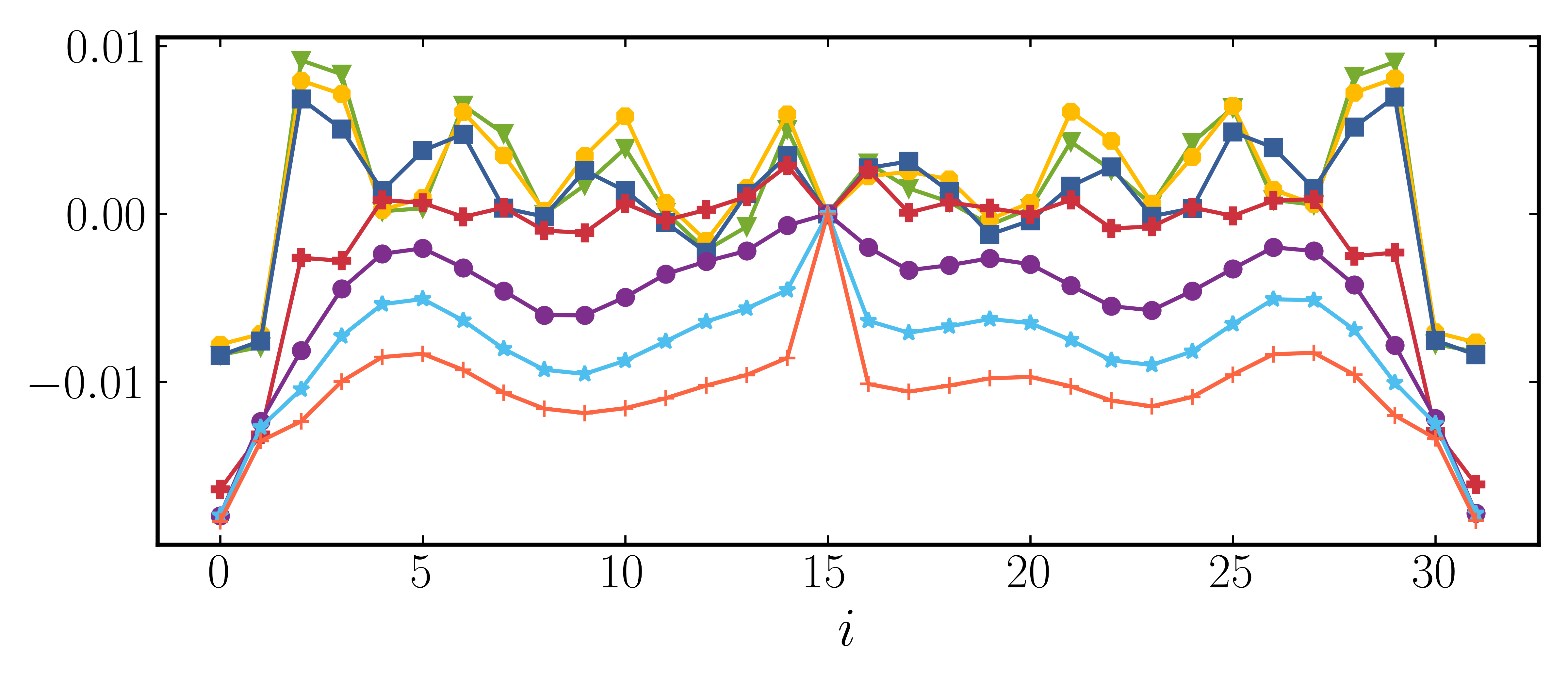}
  \caption{Spin-spin correlations for spins in the same rung,  as a function of rung position, for the case of one hole (see definition of the quantity measured in the main text). We focus on the case where the hole is at the center of the ladder studied. The results are obtained using DMRG for ladders of length 32 with $J=0.2$. Top: $1/16$ hole doping; Bottom: $1/8$ hole doping. } \label{fig:rung_szsz_hole}
\end{figure}

We then compute the ground state spin–spin correlation along rungs, $\langle S^z(i,0)n_h(L_x/2,0)S^z(i,1)\rangle$, where $(L_x/2,0)$ again denotes the position of the hole at the center of the ladder. The results for a lightly hole doped system are shown in Fig.~\ref{fig:rung_szsz_hole} (top panel), where we plot the rung correlations for different values of 
$t_2$. The correlation vanishes at the rung containing the hole, and gradually recovers to its undoped value far from the hole. The spatial range over which the correlation is suppressed provides an estimation of the polaron size \cite{my_tJ_chain}. From these results, we find that a negative $t_2$ significantly {\it enlarges} the polaron, whereas for small or positive $t_2$, the polaron remains largely confined to the middle rung. It is clear to the eye, that the negative and positive $t_2$ regions
are qualitatively different, supporting again the notion that negative $t_2$ separates spin and charge.

\begin{figure}
	\centering
  \includegraphics[width=0.5\textwidth]{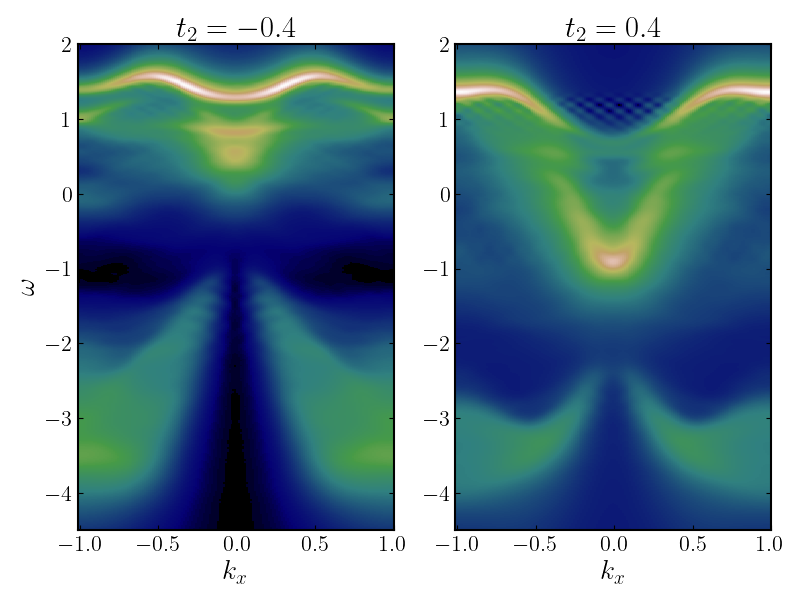} 
  \caption{Single-particle removal function, $A({\bf k},\omega)$, for $J=0.4$ at half-filling. The results are obtained using tDMRG for ladders of length $L_x=32$.} \label{fig:spectra_t2_half}
\end{figure}

The enlargement of the polaron size signals the deconfinement of the holon and the spin excitations.  
It is also a precursor to the emergence of quasi–spin–charge separation, as the ``free'' hole motion begins to dominate over the coherent quasiparticle dynamics. This leads to a suppression of the coherent peak and a redistribution of spectral weight in the photoemission spectrum. To illustrate this effect, we compute the photoemission spectra of the ladder at half filling for both positive and negative $t_2$ using tDMRG.
The results of photoemission spectrum, i.e. single-particle removal function, are displayed in Fig.~\ref{fig:spectra_t2_half}. 
As $t_2$ changes from positive to negative,
the low-energy polaron dispersion evolves from a simple cosine-like form into a band with two local minima, indicating the emergence of effective longer-range (second-order) hopping processes mediated by the spin background.
Moreover, a continuum emerges from the lowest-energy peak and becomes separated from it by a small gap. This indicates the coexistence of SCS-like behavior and hole motion in the incoherent spin background \cite{Adrian_finite_temperature}.

\begin{figure}
	\centering
  \includegraphics[width=0.4\textwidth]{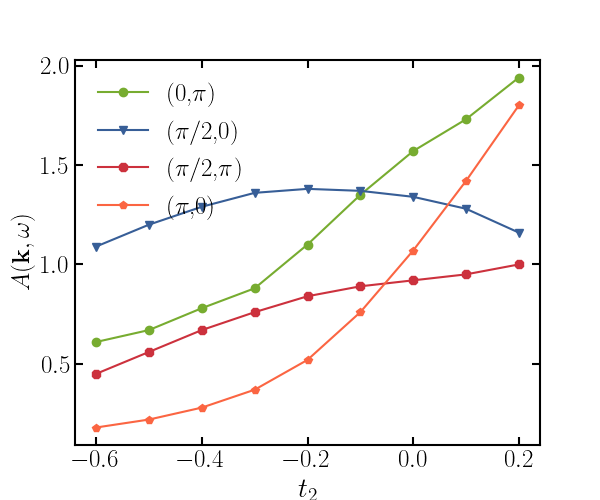} 
  \caption{The height of the first excitation peak in the photoemission spectrum as a function of $t_2$ at different momenta. The results are obtained using tDMRG for ladders of length $L_x=16$ at half-filling. Note that for some momenta the value decreases substantially
as $t_2$ becomes more negative, while for one momentum $(\pi/2,0)$ the values are approximately constant, indicating that the effect
we are investigating is momentum dependent.} \label{fig:spectra_cut_half_filling}
\end{figure}

We extract the lowest energy peak of the single-particle removal spectrum $A(\omega)$ at several momenta ${\bf k}$, and plot it as a function of $t_2$. The results are shown in Fig.~\ref{fig:spectra_cut_half_filling}. We find the low-energy spectral weight at $(\pi,0)$, $(0,\pi)$, and $(\pi/2,\pi)$ are all decreasing substantially as $t_2$ becomes negative. The reduction of the quasiparticle weight $Z_{\bf K}$, linked one-to-one with the peak height, is typically associated with spin and charge separation. However, we also observe that the spin-charge separation behavior is momentum dependent. Opposite to the other cases, at momentum $(\pi/2,0)$ the height of the first excitation peak remains nearly unchanged as $t_2$ varies. This momentum-dependent behavior is consistent with previous studies in similar settings and using different many-body techniques \cite{Elbio_ladder,w23h-dhrk}.

It is worth noting that the peaks of our focus do not necessarily correspond to well-defined quasiparticle weights, but instead they could be a combination of both coherent and incoherent excitations. Nevertheless, the clear suppression of the magnitudes of the peaks indicates the diminishing of the associated quasi-particle characters.


\begin{figure}
	\centering
  \includegraphics[width=0.5\textwidth]{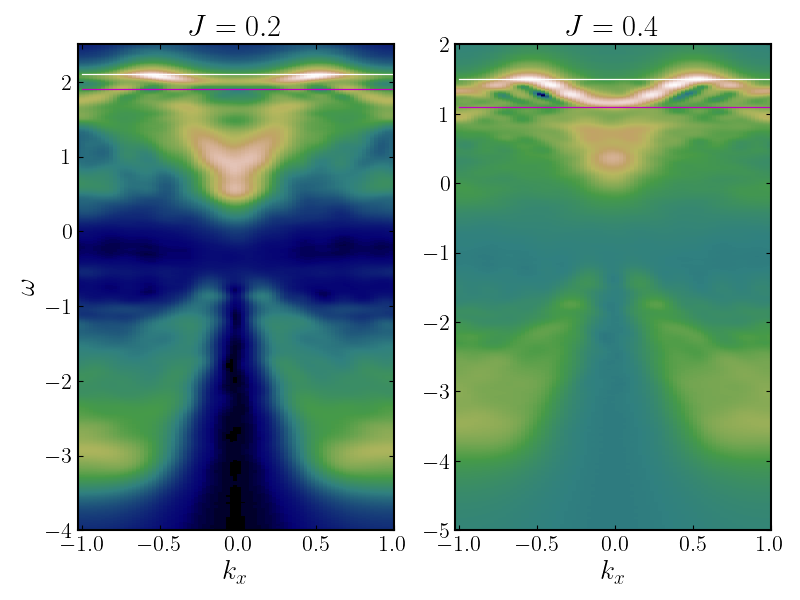} 
  \caption{Single-particle removal function for $t_2=-0.3$ at half-filling. The results are obtained using tDMRG for ladders of length $L_x=16$. The white and magenta lines are representing Fermi surface and $J$ respectively. } \label{fig:spectra_J}
\end{figure}

Our results indicate that the energy of the polaron is bounded by the spin exchange scale $J$ and remains within an energy window of order $J$ below the Fermi level. As shown in Fig.~\ref{fig:spectra_J}, decreasing $J$ causes the avoided level crossing to shift and cut through the polaron band. This also indicates that the low-energy dispersion arises from confined excitations.

\begin{figure}
	\centering
  \includegraphics[width=0.232\textwidth]{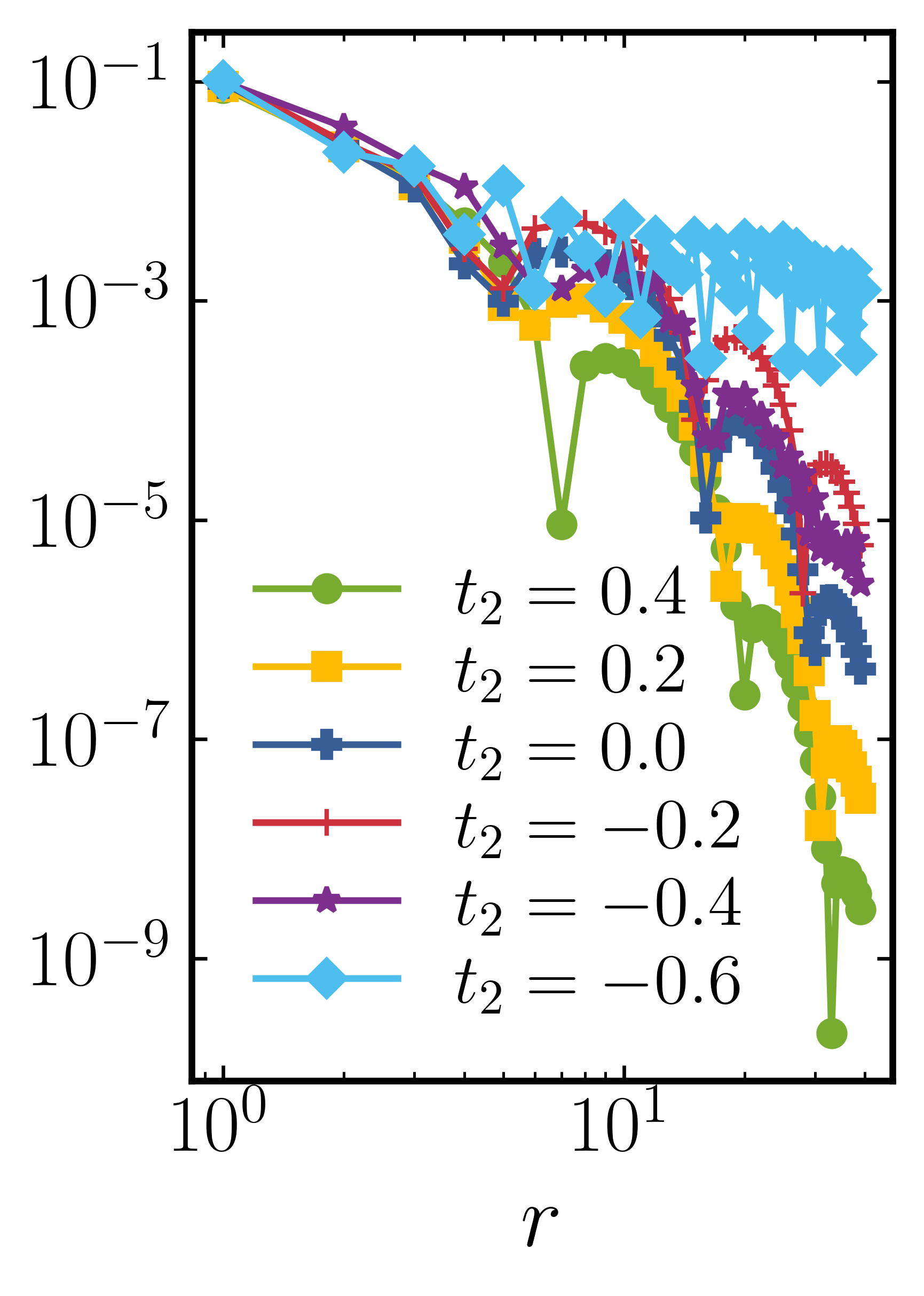} \includegraphics[width=0.244\textwidth]{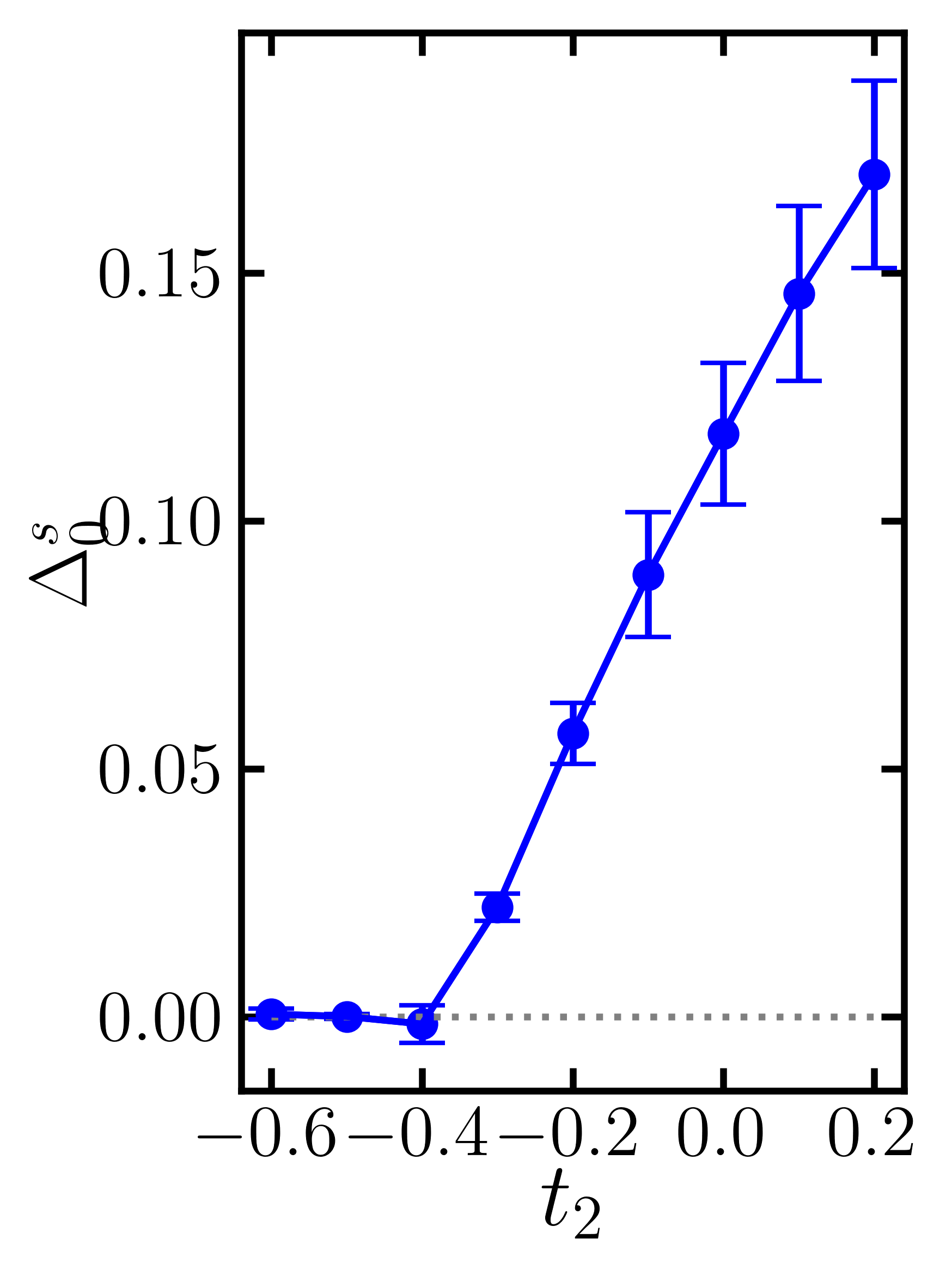} 	
  \caption{Left: Spin-spin correlations for the ladder with doping density $\delta_n = 0.0625$ with various values of $t_2$.
  Right: Spin gap extrapolated to the thermodynamic limit for the doped $t_1-t_2-J$ chain upon $1/16$ hole doping with varying $t_2$. The results are obtained using DMRG for ladders of lengths from 16 to 64, and they indicate a clear transition from spin gapped to spin gapless states.} \label{fig:spin_gap}
\end{figure}

\section{Effects of Hole Doping}

\begin{figure*}
	\centering
  \includegraphics[width=0.9\textwidth]{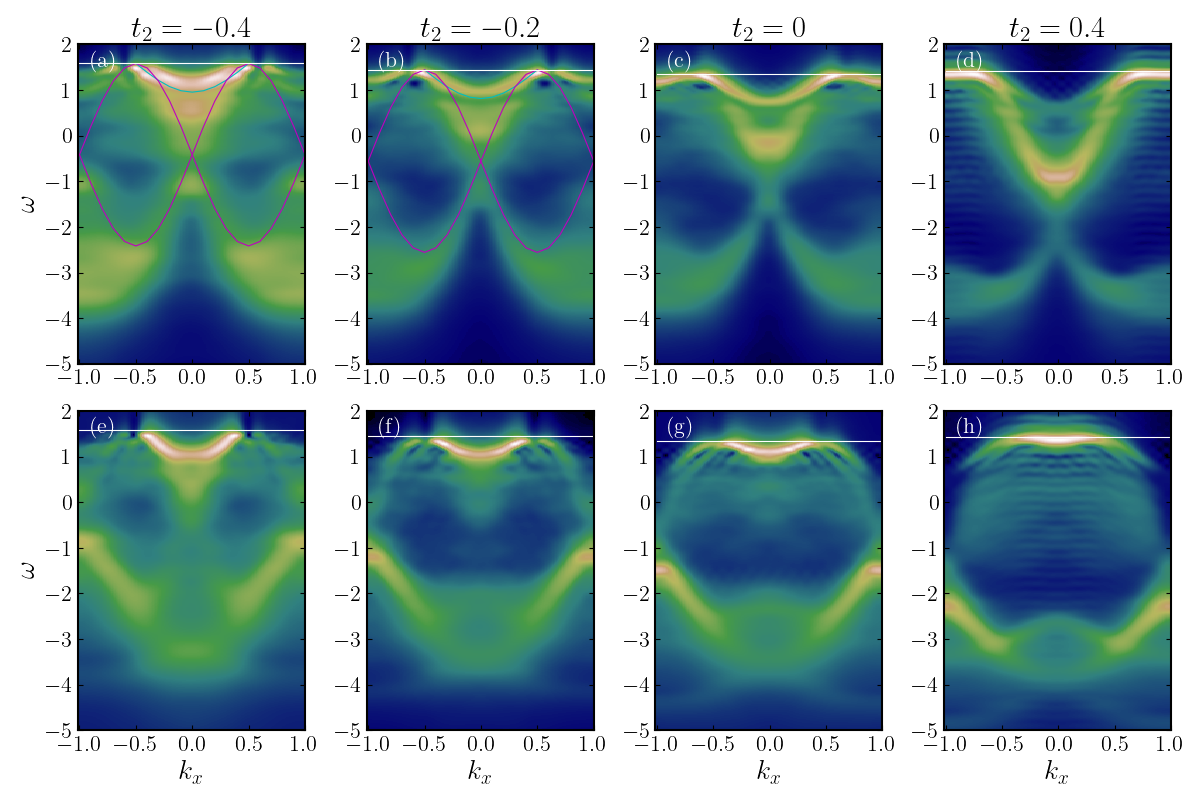} 	
  \caption{Photoemission spectrum for various values of $t_2$ with $1/16$ hole doping. The results are obtained using tDMRG for ladders of length $L_x=32$ and $J=0.4$.} \label{fig:spectra_doped}
\end{figure*}

Hole doping introduces richer physics to two-leg ladders~\cite{elbio1992,elbio1996}. The presence of doped holes weakens the rung-singlet bonds, thus enhances charge and spin dynamics along the legs. In the presence of next-nearest-neighbor (NNN) hopping, both hole doping and diagonal hopping promote magnetic excitations and charge motion along the leg direction \cite{yang2026effects}.

The enhanced mobility is closely related to the ground-state properties. In Ref.~\cite{my_3band}, one of the authors demonstrated that the three band Hubbard ladder can be mapped to a lightly doped two-leg $t-J$ ladder with $t_2/t_1\sim -0.4$ and $J/t_1\sim 0.4$, and it realizes a Luttinger liquid (LL) phase, in which both the spin and charge sectors are gapless. Although this previous study did not dive into the transition in the dynamics, one would naturally expect the characteristic properties of a Luttinger liquid to emerge, including SCS.

We first verify the transition from the rung-singlet dominated phase to a quasi-long-range ordered phase upon doping by examining ground-state correlations. In particular, we compute the spin-spin correlation function $\langle S^z_{(0,0)} S^z_{(r,0)}\rangle$ for varying values of $t_2/t_1$ at a hole doping level of $1/16$, where $(0,0)$ denotes the reference site located at $L_x/8$ from one end of the ladder to minimize boundary effects. 
The results for different values of $t_2$ are shown in Fig.~\ref{fig:spin_gap} (left panel) on a log-log scale, which facilitates the identification of the power-law decay characteristic of quasi-long-range order. We find that as $t_2$ decreases, the correlation length keeps increasing. At $t_2=-0.6$, the correlation exhibit  algebraic decay, indicating that the system has transitioned out of the rung-singlet phase and the spin gap closes. We further support these results by calculating explicitly the spin gaps extrapolated to the thermodynamic limit, and the results are shown 
in Fig.~\ref{fig:spin_gap}
(right panel). We clearly find that the spin gap closes for $t_2 \leq -0.4$, among the most important results of this publication. The extrapolation for each $t_2$ value is shown in Appendix A. This suggests that a transition from the canonical Luther-Emery regime to the Tomonaga-Luttinger regime has taken place by moving $t_2$ into more and more negative values~\cite{others}. 

We next examine how doping affects the excitations by computing the photoemission spectra at different doping levels. The results for $32\times2$ ladders are shown in Fig.~\ref{fig:spectra_doped}.
As the doping is introduced, the spin–charge separation continuum evolves from a regime where the holon moves in a background of incoherent spin fluctuations to one where it moves in a spin-liquid background, resembling the behavior in one-dimensional chains, where spinons and holons propagate with different velocities. Comparing Fig.~\ref{fig:spectra_t2_half} (a, b) and Fig.~\ref{fig:spectra_doped} (a, d), we find that the avoided level crossing in the holon branch at $(k_x=0,k_y=0)$ disappears upon doping.
For positive $t_2$ (panels (d), (f)), the hole is bounded to the spin on the same rung (see Fig.~\ref{fig:rung_szsz_hole}), forming a composite excitation. As a result, the dispersion in the photoemission spectrum resembles that of the conventional electron band centered at $\pi$ in the bonding channel. As $t_2$ decreases (panels (b), (c)), the single cosine-like band evolves into one with two local minima, and the spin-charge separation continuum emerges at higher energies, which is separated from the low-energy polaron band. 
As $t_2$ decrease further (panel (a)), the low-energy feature between $\pi/2$ to $\pi$ fades, indicating the disappearence of the polaron mode, and spin-charge separation becomes the dominating excitations in the ladder.

This behavior is also supported by the spin-spin across a hole \cite{Elbio_ladder,my_tJ_chain} correlation shown in 
Fig.~\ref{fig:rung_szsz_hole} (bottom) for $1/8$ hole dopping. When $t_2$ enters the LL regime, the polaron extends to the entire system, indicating the breakdown of the bound states, and only the dynamics of separated holes and spins is left.
It it worth noting that, within the current numerical resolution, it remains unclear whether the spin–charge separation continuum is separated from the polaron band by a finite gap. Thus, whether the polaron band is completely absent at this parameter set ($t_2=-0.4$, $n=15/16$), or not, requires further investigation.

In the one-dimensional spin–charge separation picture, the spinon dispersion is given by $\pi J|sin(k_x-\pi/2)|/2$ for $(|k_x|\leq \pi/2)$, while the holon dispersion follows $-2t_1 cos(k_x)$. 
In the two-leg ladder systems, our numerical results show that the holon branches retain a cosine-like dispersion characteristic of one-dimensional systems.
In Fig.~\ref{fig:spectra_doped} (a), (b) we plot the spinon and holon dispersions, which coincide well with the spectral peaks or continuum edges for negative $t_2$. These results provide robust evidence that spin–charge separation in this regime arises from effectively weakly decoupled chains \cite{Adrian_hubbard_ladder}, each with doping density $\delta$, rather than from the ladder as a whole. A key distinction is that, in the case where SCS is induced by doping and NNN hopping, the avoided level crossing is absent.

\begin{figure}[ht]
	\centering
  \includegraphics[width=0.4\textwidth]{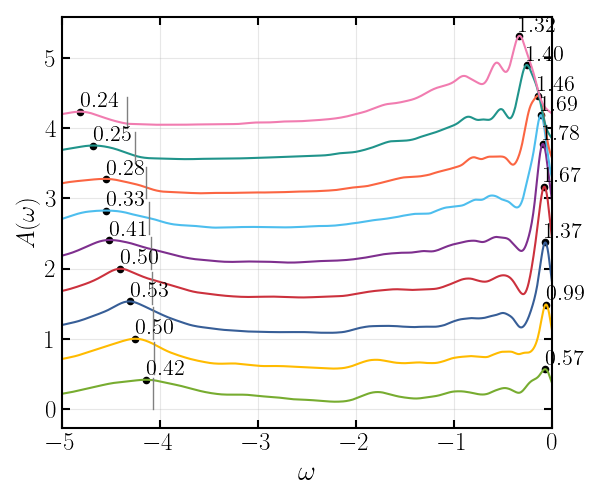} 	
  \caption{Frequency cuts of $A(\omega)$ at momentum $(\pi/2,0)$ for various values of $t_2$. The results are obtained using tDMRG for ladders of length $L_x=16$. $t_2$ decreases from 0.4 to -0.4 in steps of 0.1 for the curves shown from top to bottom. Each curve is shifted by 0.5 for clarity.} \label{fig:spectra_cut_pi_2}
\end{figure}

\begin{figure}
	\centering
  \includegraphics[width=0.4\textwidth]{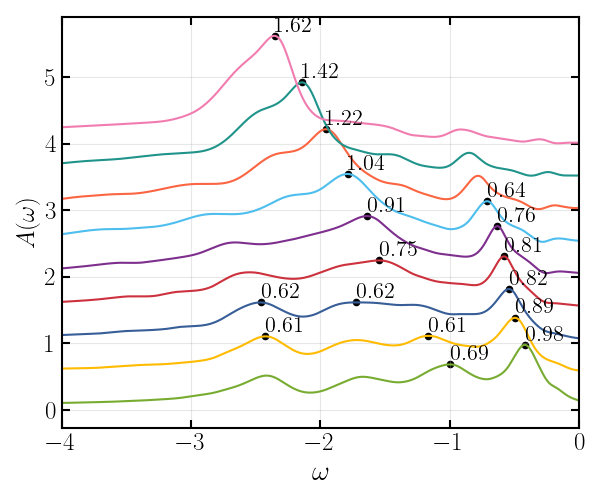} 	
  \caption{Frequency cuts of $A(\omega)$ at momentum $(0,0)$ for various values of $t_2$. The results are obtained using tDMRG for ladders of length $L_x=16$. $t_2$ decreases from 0.4 to -0.4 in steps of 0.1 for the curves shown from top to bottom. Each curve is shifted by 0.5 for clarity.} \label{fig:spectra_cut_0}
\end{figure}

To better visualize the holon branch, we plot the spectral function along the cut $k_x=\pi/2$ in the bonding channel (Fig.~\ref{fig:spectra_cut_pi_2}), which lies close to the effective Fermi momentum $k_F$ of a one-dimensional chain. If the system recovers the one-dimensional spin-charge separation picture and the first excitation peak corresponds to the top of the holon branch, the bottom of the holon branch would be at $-4t_1$. We therefore mark the position of the bottom of the holon branch (at $-4t_1$) with a short gray line, assuming that the first excitation belongs to the holon branch.  As $t_2$ decreases toward negative values, the bandwidth approaches the expected value of $4t_1$, and the spectral weight becomes more evenly distributed between the spinon and holon branches. 
The cut along $k_x=0$ in the bonding channel (Fig.~\ref{fig:spectra_cut_0}) shows that, as $t_2$ decreases, the coherent low-energy mode splits into two higher-energy modes, accompanied by a redistribution of spectral weight, which also indicates the emergence of spin-charge separation.


\section{Conclusions}

Our numerical study of the two-leg ladder demonstrates that the nature of spin correlations and spin excitations plays an important role in the emergence of spin–charge separation (SCS). For ladders with an even number of legs, the half-filled system is typically spin gapped. In this regime, hole motion is strongly coupled to the spin excitations, leading to the formation of bound states hole-spin (polarons). Consequently, it is nontrivial to identify the conditions under which one-dimensional-like physics can emerge in two-leg $t-J$ ladder systems.

By systematically investigating the effects of adding next-nearest-neighbor hopping $t_2$ with the proper sign, as well as hole doping, we find that both factors promote the separation of spin and charge excitations. In the conventional Hubbard ladder with only nearest-neighbor hopping, the system remains a spin-gapped insulator within a moderate range of hole doping, 
and the system typically realizes a Luther–Emery phase upon light doping. In contrast, the three-band Hubbard ladder has been shown to exhibit Luttinger liquid behavior upon light doping, which can be effectively described by a $t_1-t_2-J$ model \cite{my_3band}. Our results for this effective model highlight that both finite doping and negative $t_2$ are essential ingredients for the recovery of 1D-like physics.

A natural question is whether the Luttinger liquid phase is a necessary condition for SCS. Our results suggest that this is not the case. Previous studies have reported signatures of SCS in the weak-coupling limit of two-leg Hubbard ladders \cite{Adrian_hubbard_ladder}, away from the $t-J$ model regime, where a continuum emerges from the polaron band. Consistently, we find that even at half filling, a large negative $t_2$ can induce SCS upon introducing a single hole. Although an adequate spin correlation is required to support mobile spinon excitations, a fully closed spin gap is not strictly necessary.

These findings suggest that SCS can arise beyond strictly one-dimensional systems and may persist in wider ladders. 
Whether this phenomenon can exist all the way to two-dimensional systems is an intriguing and challenging question that
merits further work in the future. It would also be interesting to find if other Luther-Emery systems, as in the recently proposed 
multiorbital fermionic generalization of the Haldane chain~\cite{pontus,patel1,patel2}, also transition to a TLL by adding additional hoppings.

\section{Acknowledgment}

This work was supported by the U.S. Department of Energy (DOE), Office of Science, Basic Energy Sciences (BES), Materials Sciences and Engineering Division.

\appendix

\section{Spin gap extrapolations}

\begin{figure}
	\centering
  \includegraphics[width=0.35\textwidth]{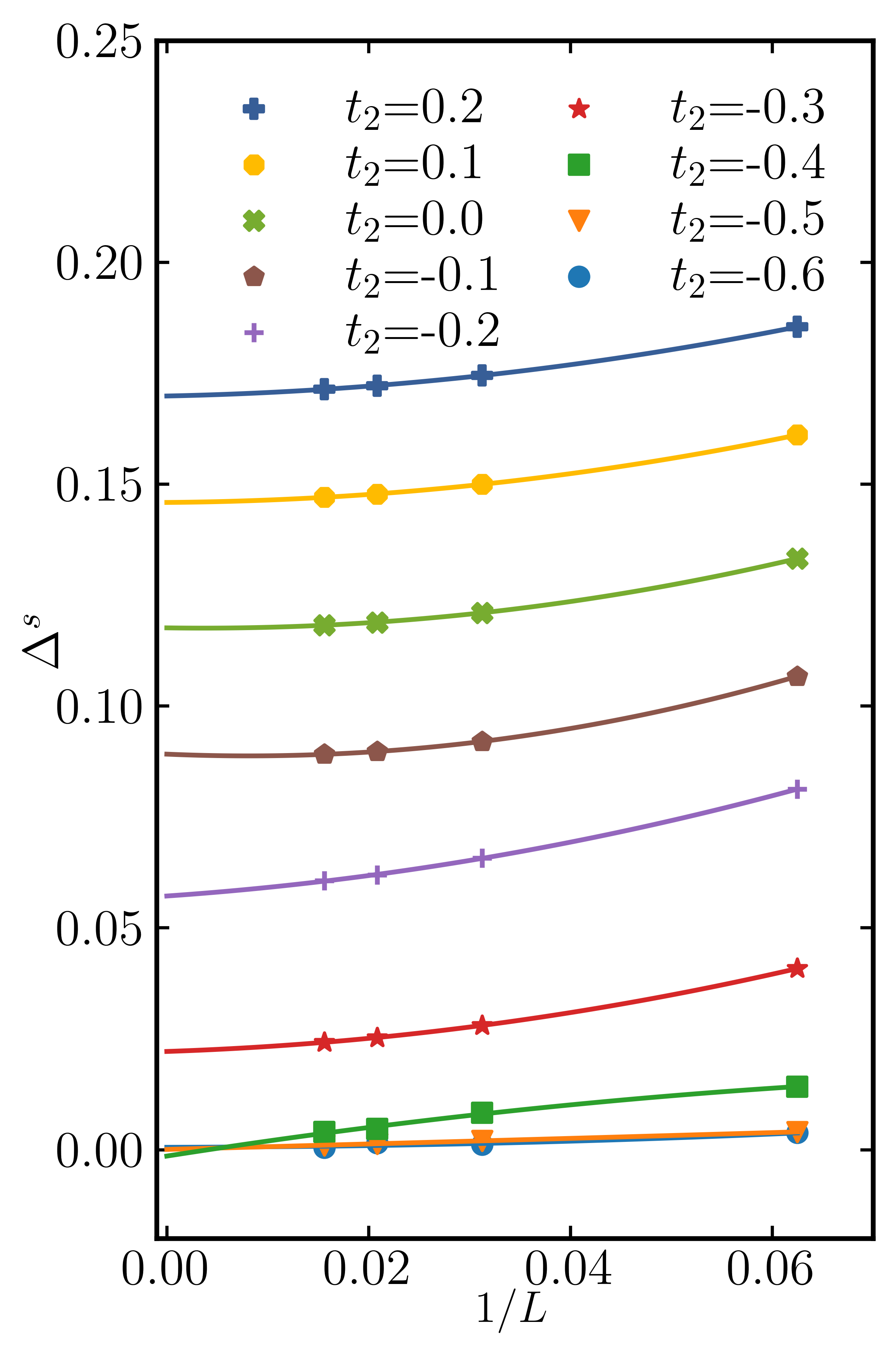} 	
  \caption{The finite-size spin gaps as a function of $1/L_x$, extrapolated
to the thermodynamic limit ($1/Lx \to 0$) using a second-order polynomial fit.} \label{fig:extrapolation_spin_gap}
\end{figure}

In Fig.~\ref{fig:extrapolation_spin_gap} we show the spin gaps extrapolations using a second-order polynomial fit. As $t_2$ decreases, the fit becomes increasingly linear, and the fitting error correspondingly decreases (see Fig.~\ref{fig:spin_gap} in main text).  

\bibliography{ref}

\begin{thebibliography}{45}%
\makeatletter
\providecommand \@ifxundefined [1]{%
 \@ifx{#1\undefined}
}%
\providecommand \@ifnum [1]{%
 \ifnum #1\expandafter \@firstoftwo
 \else \expandafter \@secondoftwo
 \fi
}%
\providecommand \@ifx [1]{%
 \ifx #1\expandafter \@firstoftwo
 \else \expandafter \@secondoftwo
 \fi
}%
\providecommand \natexlab [1]{#1}%
\providecommand \enquote  [1]{``#1''}%
\providecommand \bibnamefont  [1]{#1}%
\providecommand \bibfnamefont [1]{#1}%
\providecommand \citenamefont [1]{#1}%
\providecommand \href@noop [0]{\@secondoftwo}%
\providecommand \href [0]{\begingroup \@sanitize@url \@href}%
\providecommand \@href[1]{\@@startlink{#1}\@@href}%
\providecommand \@@href[1]{\endgroup#1\@@endlink}%
\providecommand \@sanitize@url [0]{\catcode `\\12\catcode `\$12\catcode `\&12\catcode `\#12\catcode `\^12\catcode `\_12\catcode `\%12\relax}%
\providecommand \@@startlink[1]{}%
\providecommand \@@endlink[0]{}%
\providecommand \url  [0]{\begingroup\@sanitize@url \@url }%
\providecommand \@url [1]{\endgroup\@href {#1}{\urlprefix }}%
\providecommand \urlprefix  [0]{URL }%
\providecommand \Eprint [0]{\href }%
\providecommand \doibase [0]{https://doi.org/}%
\providecommand \selectlanguage [0]{\@gobble}%
\providecommand \bibinfo  [0]{\@secondoftwo}%
\providecommand \bibfield  [0]{\@secondoftwo}%
\providecommand \translation [1]{[#1]}%
\providecommand \BibitemOpen [0]{}%
\providecommand \bibitemStop [0]{}%
\providecommand \bibitemNoStop [0]{.\EOS\space}%
\providecommand \EOS [0]{\spacefactor3000\relax}%
\providecommand \BibitemShut  [1]{\csname bibitem#1\endcsname}%
\let\auto@bib@innerbib\@empty
\bibitem [{\citenamefont {Kivelson}\ and\ \citenamefont {Rokhsar}(1990)}]{kivelson1990}%
  \BibitemOpen
  \bibfield  {author} {\bibinfo {author} {\bibfnamefont {S.~A.}\ \bibnamefont {Kivelson}}\ and\ \bibinfo {author} {\bibfnamefont {D.~S.}\ \bibnamefont {Rokhsar}},\ }\bibfield  {title} {\bibinfo {title} {Bogoliubov quasiparticles, spinons, and spin-charge decoupling in superconductors},\ }\href {https://doi.org/10.1103/PhysRevB.41.11693} {\bibfield  {journal} {\bibinfo  {journal} {Phys. Rev. B}\ }\textbf {\bibinfo {volume} {41}},\ \bibinfo {pages} {11693} (\bibinfo {year} {1990})}\BibitemShut {NoStop}%
\bibitem [{\citenamefont {Mudry}\ and\ \citenamefont {Fradkin}(1994)}]{murdy1994}%
  \BibitemOpen
  \bibfield  {author} {\bibinfo {author} {\bibfnamefont {C.}~\bibnamefont {Mudry}}\ and\ \bibinfo {author} {\bibfnamefont {E.}~\bibnamefont {Fradkin}},\ }\bibfield  {title} {\bibinfo {title} {Separation of spin and charge quantum numbers in strongly correlated systems},\ }\href {https://doi.org/10.1103/PhysRevB.49.5200} {\bibfield  {journal} {\bibinfo  {journal} {Phys. Rev. B}\ }\textbf {\bibinfo {volume} {49}},\ \bibinfo {pages} {5200} (\bibinfo {year} {1994})}\BibitemShut {NoStop}%
\bibitem [{\citenamefont {Zhao}\ and\ \citenamefont {Hershfield}(1995)}]{zhao1995}%
  \BibitemOpen
  \bibfield  {author} {\bibinfo {author} {\bibfnamefont {H.~L.}\ \bibnamefont {Zhao}}\ and\ \bibinfo {author} {\bibfnamefont {S.}~\bibnamefont {Hershfield}},\ }\bibfield  {title} {\bibinfo {title} {Tunneling, relaxation of spin-polarized quasiparticles, and spin-charge separation in superconductors},\ }\href {https://doi.org/10.1103/PhysRevB.52.3632} {\bibfield  {journal} {\bibinfo  {journal} {Phys. Rev. B}\ }\textbf {\bibinfo {volume} {52}},\ \bibinfo {pages} {3632} (\bibinfo {year} {1995})}\BibitemShut {NoStop}%
\bibitem [{\citenamefont {Si}(1997)}]{si1997}%
  \BibitemOpen
  \bibfield  {author} {\bibinfo {author} {\bibfnamefont {Q.}~\bibnamefont {Si}},\ }\bibfield  {title} {\bibinfo {title} {Spin conductivity and spin-charge separation in the high- ${T}_{c}$ cuprates},\ }\href {https://doi.org/10.1103/PhysRevLett.78.1767} {\bibfield  {journal} {\bibinfo  {journal} {Phys. Rev. Lett.}\ }\textbf {\bibinfo {volume} {78}},\ \bibinfo {pages} {1767} (\bibinfo {year} {1997})}\BibitemShut {NoStop}%
\bibitem [{\citenamefont {Si}(1998)}]{si1998}%
  \BibitemOpen
  \bibfield  {author} {\bibinfo {author} {\bibfnamefont {Q.}~\bibnamefont {Si}},\ }\bibfield  {title} {\bibinfo {title} {Spin injection into a luttinger liquid},\ }\href {https://doi.org/10.1103/PhysRevLett.81.3191} {\bibfield  {journal} {\bibinfo  {journal} {Phys. Rev. Lett.}\ }\textbf {\bibinfo {volume} {81}},\ \bibinfo {pages} {3191} (\bibinfo {year} {1998})}\BibitemShut {NoStop}%
\bibitem [{\citenamefont {Nagaosa}(1998)}]{Nagaosa1998}%
  \BibitemOpen
  \bibfield  {author} {\bibinfo {author} {\bibfnamefont {N.}~\bibnamefont {Nagaosa}},\ }\bibfield  {title} {\bibinfo {title} {Spin-charge separation in strongly correlated electronic systems},\ }\href {https://doi.org/10.1088/0953-8984/10/49/025} {\bibfield  {journal} {\bibinfo  {journal} {Journal of Physics: Condensed Matter}\ }\textbf {\bibinfo {volume} {10}},\ \bibinfo {pages} {11385} (\bibinfo {year} {1998})}\BibitemShut {NoStop}%
\bibitem [{\citenamefont {Balents}\ and\ \citenamefont {Egger}(2000)}]{balents2000}%
  \BibitemOpen
  \bibfield  {author} {\bibinfo {author} {\bibfnamefont {L.}~\bibnamefont {Balents}}\ and\ \bibinfo {author} {\bibfnamefont {R.}~\bibnamefont {Egger}},\ }\bibfield  {title} {\bibinfo {title} {Spin transport in interacting quantum wires and carbon nanotubes},\ }\href {https://doi.org/10.1103/PhysRevLett.85.3464} {\bibfield  {journal} {\bibinfo  {journal} {Phys. Rev. Lett.}\ }\textbf {\bibinfo {volume} {85}},\ \bibinfo {pages} {3464} (\bibinfo {year} {2000})}\BibitemShut {NoStop}%
\bibitem [{\citenamefont {Kim}\ \emph {et~al.}(1996)\citenamefont {Kim}, \citenamefont {Matsuura}, \citenamefont {Shen}, \citenamefont {Motoyama}, \citenamefont {Eisaki}, \citenamefont {Uchida}, \citenamefont {Tohyama},\ and\ \citenamefont {Maekawa}}]{kim1996observation}%
  \BibitemOpen
  \bibfield  {author} {\bibinfo {author} {\bibfnamefont {C.}~\bibnamefont {Kim}}, \bibinfo {author} {\bibfnamefont {A.~Y.}\ \bibnamefont {Matsuura}}, \bibinfo {author} {\bibfnamefont {Z.-X.}\ \bibnamefont {Shen}}, \bibinfo {author} {\bibfnamefont {N.}~\bibnamefont {Motoyama}}, \bibinfo {author} {\bibfnamefont {H.}~\bibnamefont {Eisaki}}, \bibinfo {author} {\bibfnamefont {S.}~\bibnamefont {Uchida}}, \bibinfo {author} {\bibfnamefont {T.}~\bibnamefont {Tohyama}},\ and\ \bibinfo {author} {\bibfnamefont {S.}~\bibnamefont {Maekawa}},\ }\bibfield  {title} {\bibinfo {title} {Observation of spin-charge separation in one-dimensional ${S}r{C}u{O}_{2}$},\ }\href {https://doi.org/10.1103/PhysRevLett.77.4054} {\bibfield  {journal} {\bibinfo  {journal} {Phys. Rev. Lett.}\ }\textbf {\bibinfo {volume} {77}},\ \bibinfo {pages} {4054} (\bibinfo {year} {1996})}\BibitemShut {NoStop}%
\bibitem [{\citenamefont {Sing}\ \emph {et~al.}(2003)\citenamefont {Sing}, \citenamefont {Schwingenschl\"ogl}, \citenamefont {Claessen}, \citenamefont {Blaha}, \citenamefont {Carmelo}, \citenamefont {Martelo}, \citenamefont {Sacramento}, \citenamefont {Dressel},\ and\ \citenamefont {Jacobsen}}]{sing2003}%
  \BibitemOpen
  \bibfield  {author} {\bibinfo {author} {\bibfnamefont {M.}~\bibnamefont {Sing}}, \bibinfo {author} {\bibfnamefont {U.}~\bibnamefont {Schwingenschl\"ogl}}, \bibinfo {author} {\bibfnamefont {R.}~\bibnamefont {Claessen}}, \bibinfo {author} {\bibfnamefont {P.}~\bibnamefont {Blaha}}, \bibinfo {author} {\bibfnamefont {J.~M.~P.}\ \bibnamefont {Carmelo}}, \bibinfo {author} {\bibfnamefont {L.~M.}\ \bibnamefont {Martelo}}, \bibinfo {author} {\bibfnamefont {P.~D.}\ \bibnamefont {Sacramento}}, \bibinfo {author} {\bibfnamefont {M.}~\bibnamefont {Dressel}},\ and\ \bibinfo {author} {\bibfnamefont {C.~S.}\ \bibnamefont {Jacobsen}},\ }\bibfield  {title} {\bibinfo {title} {Electronic structure of the quasi-one-dimensional organic conductor {TTF-TCNQ}},\ }\href {https://doi.org/10.1103/PhysRevB.68.125111} {\bibfield  {journal} {\bibinfo  {journal} {Phys. Rev. B}\ }\textbf {\bibinfo {volume} {68}},\ \bibinfo {pages} {125111} (\bibinfo {year} {2003})}\BibitemShut {NoStop}%
\bibitem [{\citenamefont {Auslaender}\ \emph {et~al.}(2005)\citenamefont {Auslaender}, \citenamefont {Steinberg}, \citenamefont {Yacoby}, \citenamefont {Tserkovnyak}, \citenamefont {Halperin}, \citenamefont {Baldwin}, \citenamefont {Pfeiffer},\ and\ \citenamefont {West}}]{auslaender2005spin}%
  \BibitemOpen
  \bibfield  {author} {\bibinfo {author} {\bibfnamefont {O.}~\bibnamefont {Auslaender}}, \bibinfo {author} {\bibfnamefont {H.}~\bibnamefont {Steinberg}}, \bibinfo {author} {\bibfnamefont {A.}~\bibnamefont {Yacoby}}, \bibinfo {author} {\bibfnamefont {Y.}~\bibnamefont {Tserkovnyak}}, \bibinfo {author} {\bibfnamefont {B.}~\bibnamefont {Halperin}}, \bibinfo {author} {\bibfnamefont {K.}~\bibnamefont {Baldwin}}, \bibinfo {author} {\bibfnamefont {L.}~\bibnamefont {Pfeiffer}},\ and\ \bibinfo {author} {\bibfnamefont {K.}~\bibnamefont {West}},\ }\bibfield  {title} {\bibinfo {title} {Spin-charge separation and localization in one dimension},\ }\href@noop {} {\bibfield  {journal} {\bibinfo  {journal} {Science}\ }\textbf {\bibinfo {volume} {308}},\ \bibinfo {pages} {88} (\bibinfo {year} {2005})}\BibitemShut {NoStop}%
\bibitem [{\citenamefont {Kim}\ \emph {et~al.}(2006)\citenamefont {Kim}, \citenamefont {Koh}, \citenamefont {Rotenberg}, \citenamefont {Oh}, \citenamefont {Eisaki}, \citenamefont {Motoyama}, \citenamefont {Uchida}, \citenamefont {Tohyama}, \citenamefont {Maekawa}, \citenamefont {Shen} \emph {et~al.}}]{kim2006distinct}%
  \BibitemOpen
  \bibfield  {author} {\bibinfo {author} {\bibfnamefont {B.}~\bibnamefont {Kim}}, \bibinfo {author} {\bibfnamefont {H.}~\bibnamefont {Koh}}, \bibinfo {author} {\bibfnamefont {E.}~\bibnamefont {Rotenberg}}, \bibinfo {author} {\bibfnamefont {S.-J.}\ \bibnamefont {Oh}}, \bibinfo {author} {\bibfnamefont {H.}~\bibnamefont {Eisaki}}, \bibinfo {author} {\bibfnamefont {N.}~\bibnamefont {Motoyama}}, \bibinfo {author} {\bibfnamefont {S.-i.}\ \bibnamefont {Uchida}}, \bibinfo {author} {\bibfnamefont {T.}~\bibnamefont {Tohyama}}, \bibinfo {author} {\bibfnamefont {S.}~\bibnamefont {Maekawa}}, \bibinfo {author} {\bibfnamefont {Z.-X.}\ \bibnamefont {Shen}}, \emph {et~al.},\ }\bibfield  {title} {\bibinfo {title} {Distinct spinon and holon dispersions in photoemission spectral functions from one-dimensional ${S}r{C}u{O}_2$},\ }\href@noop {} {\bibfield  {journal} {\bibinfo  {journal} {Nature Physics}\ }\textbf {\bibinfo {volume} {2}},\ \bibinfo {pages} {397} (\bibinfo {year} {2006})}\BibitemShut {NoStop}%
\bibitem [{\citenamefont {Jompol}\ \emph {et~al.}(2009)\citenamefont {Jompol}, \citenamefont {Ford}, \citenamefont {Griffiths}, \citenamefont {Farrer}, \citenamefont {Jones}, \citenamefont {Anderson}, \citenamefont {Ritchie}, \citenamefont {Silk},\ and\ \citenamefont {Schofield}}]{jompol2009probing}%
  \BibitemOpen
  \bibfield  {author} {\bibinfo {author} {\bibfnamefont {Y.}~\bibnamefont {Jompol}}, \bibinfo {author} {\bibfnamefont {C.}~\bibnamefont {Ford}}, \bibinfo {author} {\bibfnamefont {J.}~\bibnamefont {Griffiths}}, \bibinfo {author} {\bibfnamefont {I.}~\bibnamefont {Farrer}}, \bibinfo {author} {\bibfnamefont {G.}~\bibnamefont {Jones}}, \bibinfo {author} {\bibfnamefont {D.}~\bibnamefont {Anderson}}, \bibinfo {author} {\bibfnamefont {D.}~\bibnamefont {Ritchie}}, \bibinfo {author} {\bibfnamefont {T.}~\bibnamefont {Silk}},\ and\ \bibinfo {author} {\bibfnamefont {A.}~\bibnamefont {Schofield}},\ }\bibfield  {title} {\bibinfo {title} {Probing spin-charge separation in a tomonaga-luttinger liquid},\ }\href@noop {} {\bibfield  {journal} {\bibinfo  {journal} {Science}\ }\textbf {\bibinfo {volume} {325}},\ \bibinfo {pages} {597} (\bibinfo {year} {2009})}\BibitemShut {NoStop}%
\bibitem [{\citenamefont {Martins}\ \emph {et~al.}(2000{\natexlab{a}})\citenamefont {Martins}, \citenamefont {Xavier}, \citenamefont {Gazza}, \citenamefont {Vojta},\ and\ \citenamefont {Dagotto}}]{Elbio_ladder}%
  \BibitemOpen
  \bibfield  {author} {\bibinfo {author} {\bibfnamefont {G.~B.}\ \bibnamefont {Martins}}, \bibinfo {author} {\bibfnamefont {J.~C.}\ \bibnamefont {Xavier}}, \bibinfo {author} {\bibfnamefont {C.}~\bibnamefont {Gazza}}, \bibinfo {author} {\bibfnamefont {M.}~\bibnamefont {Vojta}},\ and\ \bibinfo {author} {\bibfnamefont {E.}~\bibnamefont {Dagotto}},\ }\bibfield  {title} {\bibinfo {title} {Indications of spin-charge separation at short distance and stripe formation in the extended t-j model on ladders and planes},\ }\href {https://doi.org/10.1103/PhysRevB.63.014414} {\bibfield  {journal} {\bibinfo  {journal} {Phys. Rev. B}\ }\textbf {\bibinfo {volume} {63}},\ \bibinfo {pages} {014414} (\bibinfo {year} {2000}{\natexlab{a}})}\BibitemShut {NoStop}%
\bibitem [{\citenamefont {Martins}\ \emph {et~al.}(1999)\citenamefont {Martins}, \citenamefont {Eder},\ and\ \citenamefont {Dagotto}}]{elbio_2D}%
  \BibitemOpen
  \bibfield  {author} {\bibinfo {author} {\bibfnamefont {G.~B.}\ \bibnamefont {Martins}}, \bibinfo {author} {\bibfnamefont {R.}~\bibnamefont {Eder}},\ and\ \bibinfo {author} {\bibfnamefont {E.}~\bibnamefont {Dagotto}},\ }\bibfield  {title} {\bibinfo {title} {Indications of spin-charge separation in the two-dimensional extended t-j model},\ }\href {https://doi.org/10.1103/PhysRevB.60.R3716} {\bibfield  {journal} {\bibinfo  {journal} {Phys. Rev. B}\ }\textbf {\bibinfo {volume} {60}},\ \bibinfo {pages} {R3716} (\bibinfo {year} {1999})}\BibitemShut {NoStop}%
\bibitem [{\citenamefont {Tohyama}\ \emph {et~al.}(1999)\citenamefont {Tohyama}, \citenamefont {Gazza}, \citenamefont {Shih}, \citenamefont {Chen}, \citenamefont {Lee}, \citenamefont {Maekawa},\ and\ \citenamefont {Dagotto}}]{elbio1999}%
  \BibitemOpen
  \bibfield  {author} {\bibinfo {author} {\bibfnamefont {T.}~\bibnamefont {Tohyama}}, \bibinfo {author} {\bibfnamefont {C.}~\bibnamefont {Gazza}}, \bibinfo {author} {\bibfnamefont {C.~T.}\ \bibnamefont {Shih}}, \bibinfo {author} {\bibfnamefont {Y.~C.}\ \bibnamefont {Chen}}, \bibinfo {author} {\bibfnamefont {T.~K.}\ \bibnamefont {Lee}}, \bibinfo {author} {\bibfnamefont {S.}~\bibnamefont {Maekawa}},\ and\ \bibinfo {author} {\bibfnamefont {E.}~\bibnamefont {Dagotto}},\ }\bibfield  {title} {\bibinfo {title} {Stripe stability in the extended $t\ensuremath{-}{J}$ model on planes and four-leg ladders},\ }\href {https://doi.org/10.1103/PhysRevB.59.R11649} {\bibfield  {journal} {\bibinfo  {journal} {Phys. Rev. B}\ }\textbf {\bibinfo {volume} {59}},\ \bibinfo {pages} {R11649} (\bibinfo {year} {1999})}\BibitemShut {NoStop}%
\bibitem [{\citenamefont {Martins}\ \emph {et~al.}(2000{\natexlab{b}})\citenamefont {Martins}, \citenamefont {Gazza}, \citenamefont {Xavier}, \citenamefont {Feiguin},\ and\ \citenamefont {Dagotto}}]{elbio2000}%
  \BibitemOpen
  \bibfield  {author} {\bibinfo {author} {\bibfnamefont {G.~B.}\ \bibnamefont {Martins}}, \bibinfo {author} {\bibfnamefont {C.}~\bibnamefont {Gazza}}, \bibinfo {author} {\bibfnamefont {J.~C.}\ \bibnamefont {Xavier}}, \bibinfo {author} {\bibfnamefont {A.}~\bibnamefont {Feiguin}},\ and\ \bibinfo {author} {\bibfnamefont {E.}~\bibnamefont {Dagotto}},\ }\bibfield  {title} {\bibinfo {title} {Doped stripes in models for the cuprates emerging from the one-hole properties of the insulator},\ }\href {https://doi.org/10.1103/PhysRevLett.84.5844} {\bibfield  {journal} {\bibinfo  {journal} {Phys. Rev. Lett.}\ }\textbf {\bibinfo {volume} {84}},\ \bibinfo {pages} {5844} (\bibinfo {year} {2000}{\natexlab{b}})}\BibitemShut {NoStop}%
\bibitem [{\citenamefont {Nyhegn}\ \emph {et~al.}(2025)\citenamefont {Nyhegn}, \citenamefont {Nielsen}, \citenamefont {Balents},\ and\ \citenamefont {Bruun}}]{w23h-dhrk}%
  \BibitemOpen
  \bibfield  {author} {\bibinfo {author} {\bibfnamefont {J.~H.}\ \bibnamefont {Nyhegn}}, \bibinfo {author} {\bibfnamefont {K.~K.}\ \bibnamefont {Nielsen}}, \bibinfo {author} {\bibfnamefont {L.}~\bibnamefont {Balents}},\ and\ \bibinfo {author} {\bibfnamefont {G.~M.}\ \bibnamefont {Bruun}},\ }\bibfield  {title} {\bibinfo {title} {Spin-charge bound states and emerging fermions in a quantum spin liquid},\ }\href {https://doi.org/10.1103/w23h-dhrk} {\bibfield  {journal} {\bibinfo  {journal} {PRX Quantum}\ }\textbf {\bibinfo {volume} {6}},\ \bibinfo {pages} {040347} (\bibinfo {year} {2025})}\BibitemShut {NoStop}%
\bibitem [{\citenamefont {Hyun}\ \emph {et~al.}(2025)\citenamefont {Hyun}, \citenamefont {Lee}, \citenamefont {Lim}, \citenamefont {Lee}, \citenamefont {Cha}, \citenamefont {Ahn}, \citenamefont {Jho}, \citenamefont {Gim}, \citenamefont {Park}, \citenamefont {Hashimoto}, \citenamefont {Lu}, \citenamefont {Kim},\ and\ \citenamefont {Kim}}]{hyun2025}%
  \BibitemOpen
  \bibfield  {author} {\bibinfo {author} {\bibfnamefont {J.}~\bibnamefont {Hyun}}, \bibinfo {author} {\bibfnamefont {Y.}~\bibnamefont {Lee}}, \bibinfo {author} {\bibfnamefont {C.-y.}\ \bibnamefont {Lim}}, \bibinfo {author} {\bibfnamefont {G.}~\bibnamefont {Lee}}, \bibinfo {author} {\bibfnamefont {J.}~\bibnamefont {Cha}}, \bibinfo {author} {\bibfnamefont {Y.}~\bibnamefont {Ahn}}, \bibinfo {author} {\bibfnamefont {M.}~\bibnamefont {Jho}}, \bibinfo {author} {\bibfnamefont {S.}~\bibnamefont {Gim}}, \bibinfo {author} {\bibfnamefont {M.}~\bibnamefont {Park}}, \bibinfo {author} {\bibfnamefont {M.}~\bibnamefont {Hashimoto}}, \bibinfo {author} {\bibfnamefont {D.}~\bibnamefont {Lu}}, \bibinfo {author} {\bibfnamefont {Y.}~\bibnamefont {Kim}},\ and\ \bibinfo {author} {\bibfnamefont {S.}~\bibnamefont {Kim}},\ }\bibfield  {title} {\bibinfo {title} {Band-selective spin-charge separation across the charge density wave transition in quasi-1d ${NbSe}_{3}$},\ }\href {https://doi.org/10.1103/PhysRevLett.134.206402} {\bibfield
  {journal} {\bibinfo  {journal} {Phys. Rev. Lett.}\ }\textbf {\bibinfo {volume} {134}},\ \bibinfo {pages} {206402} (\bibinfo {year} {2025})}\BibitemShut {NoStop}%
\bibitem [{\citenamefont {Linder}\ and\ \citenamefont {Robinson}(2015)}]{linder2015superconducting}%
  \BibitemOpen
  \bibfield  {author} {\bibinfo {author} {\bibfnamefont {J.}~\bibnamefont {Linder}}\ and\ \bibinfo {author} {\bibfnamefont {J.~W.}\ \bibnamefont {Robinson}},\ }\bibfield  {title} {\bibinfo {title} {Superconducting spintronics},\ }\href@noop {} {\bibfield  {journal} {\bibinfo  {journal} {Nature Physics}\ }\textbf {\bibinfo {volume} {11}},\ \bibinfo {pages} {307} (\bibinfo {year} {2015})}\BibitemShut {NoStop}%
\bibitem [{\citenamefont {Rossi}\ \emph {et~al.}(2025)\citenamefont {Rossi}, \citenamefont {Thomas}, \citenamefont {K{\"u}chle}, \citenamefont {Barr{\'e}}, \citenamefont {Yu}, \citenamefont {Zhou}, \citenamefont {Kumari}, \citenamefont {Tsai}, \citenamefont {Wong}, \citenamefont {Jozwiak} \emph {et~al.}}]{rossi2025graphene}%
  \BibitemOpen
  \bibfield  {author} {\bibinfo {author} {\bibfnamefont {A.}~\bibnamefont {Rossi}}, \bibinfo {author} {\bibfnamefont {J.~C.}\ \bibnamefont {Thomas}}, \bibinfo {author} {\bibfnamefont {J.~T.}\ \bibnamefont {K{\"u}chle}}, \bibinfo {author} {\bibfnamefont {E.}~\bibnamefont {Barr{\'e}}}, \bibinfo {author} {\bibfnamefont {Z.}~\bibnamefont {Yu}}, \bibinfo {author} {\bibfnamefont {D.}~\bibnamefont {Zhou}}, \bibinfo {author} {\bibfnamefont {S.}~\bibnamefont {Kumari}}, \bibinfo {author} {\bibfnamefont {H.-Z.}\ \bibnamefont {Tsai}}, \bibinfo {author} {\bibfnamefont {E.}~\bibnamefont {Wong}}, \bibinfo {author} {\bibfnamefont {C.}~\bibnamefont {Jozwiak}}, \emph {et~al.},\ }\bibfield  {title} {\bibinfo {title} {Graphene-driven correlated electronic states in one dimensional defects within ${WS_2}$},\ }\href@noop {} {\bibfield  {journal} {\bibinfo  {journal} {Nature communications}\ }\textbf {\bibinfo {volume} {16}},\ \bibinfo {pages} {5809} (\bibinfo {year} {2025})}\BibitemShut {NoStop}%
\bibitem [{\citenamefont {Senaratne}\ \emph {et~al.}(2022)\citenamefont {Senaratne}, \citenamefont {Cavazos-Cavazos}, \citenamefont {Wang}, \citenamefont {He}, \citenamefont {Chang}, \citenamefont {Kafle}, \citenamefont {Pu}, \citenamefont {Guan},\ and\ \citenamefont {Hulet}}]{senaratne2022spin}%
  \BibitemOpen
  \bibfield  {author} {\bibinfo {author} {\bibfnamefont {R.}~\bibnamefont {Senaratne}}, \bibinfo {author} {\bibfnamefont {D.}~\bibnamefont {Cavazos-Cavazos}}, \bibinfo {author} {\bibfnamefont {S.}~\bibnamefont {Wang}}, \bibinfo {author} {\bibfnamefont {F.}~\bibnamefont {He}}, \bibinfo {author} {\bibfnamefont {Y.-T.}\ \bibnamefont {Chang}}, \bibinfo {author} {\bibfnamefont {A.}~\bibnamefont {Kafle}}, \bibinfo {author} {\bibfnamefont {H.}~\bibnamefont {Pu}}, \bibinfo {author} {\bibfnamefont {X.-W.}\ \bibnamefont {Guan}},\ and\ \bibinfo {author} {\bibfnamefont {R.~G.}\ \bibnamefont {Hulet}},\ }\bibfield  {title} {\bibinfo {title} {Spin-charge separation in a one-dimensional fermi gas with tunable interactions},\ }\href@noop {} {\bibfield  {journal} {\bibinfo  {journal} {Science}\ }\textbf {\bibinfo {volume} {376}},\ \bibinfo {pages} {1305} (\bibinfo {year} {2022})}\BibitemShut {NoStop}%
\bibitem [{\citenamefont {Giamarchi}(2003)}]{giamarchi2003quantum}%
  \BibitemOpen
  \bibfield  {author} {\bibinfo {author} {\bibfnamefont {T.}~\bibnamefont {Giamarchi}},\ }\href@noop {} {\emph {\bibinfo {title} {Quantum physics in one dimension}}},\ Vol.\ \bibinfo {volume} {121}\ (\bibinfo  {publisher} {Clarendon press},\ \bibinfo {year} {2003})\BibitemShut {NoStop}%
\bibitem [{\citenamefont {Dagotto}\ \emph {et~al.}(1992)\citenamefont {Dagotto}, \citenamefont {Riera},\ and\ \citenamefont {Scalapino}}]{elbio1992}%
  \BibitemOpen
  \bibfield  {author} {\bibinfo {author} {\bibfnamefont {E.}~\bibnamefont {Dagotto}}, \bibinfo {author} {\bibfnamefont {J.}~\bibnamefont {Riera}},\ and\ \bibinfo {author} {\bibfnamefont {D.}~\bibnamefont {Scalapino}},\ }\bibfield  {title} {\bibinfo {title} {Superconductivity in ladders and coupled planes},\ }\href {https://doi.org/10.1103/PhysRevB.45.5744} {\bibfield  {journal} {\bibinfo  {journal} {Phys. Rev. B}\ }\textbf {\bibinfo {volume} {45}},\ \bibinfo {pages} {5744} (\bibinfo {year} {1992})}\BibitemShut {NoStop}%
\bibitem [{\citenamefont {Dagotto}\ and\ \citenamefont {Rice}(1996)}]{elbio1996}%
  \BibitemOpen
  \bibfield  {author} {\bibinfo {author} {\bibfnamefont {E.}~\bibnamefont {Dagotto}}\ and\ \bibinfo {author} {\bibfnamefont {T.}~\bibnamefont {Rice}},\ }\bibfield  {title} {\bibinfo {title} {Surprises on the way from one-to two-dimensional quantum magnets: The ladder materials},\ }\href@noop {} {\bibfield  {journal} {\bibinfo  {journal} {Science}\ }\textbf {\bibinfo {volume} {271}},\ \bibinfo {pages} {618} (\bibinfo {year} {1996})}\BibitemShut {NoStop}%
\bibitem [{\citenamefont {Dagotto}(1999)}]{elbio_1999_review}%
  \BibitemOpen
  \bibfield  {author} {\bibinfo {author} {\bibfnamefont {E.}~\bibnamefont {Dagotto}},\ }\bibfield  {title} {\bibinfo {title} {Experiments on ladders reveal a complex interplay between a spin-gapped normal state and superconductivity},\ }\href {https://doi.org/10.1088/0034-4885/62/11/202} {\bibfield  {journal} {\bibinfo  {journal} {Reports on Progress in Physics}\ }\textbf {\bibinfo {volume} {62}},\ \bibinfo {pages} {1525} (\bibinfo {year} {1999})}\BibitemShut {NoStop}%
\bibitem [{\citenamefont {L\"auchli}\ and\ \citenamefont {Poilblanc}(2004)}]{lauchili2004}%
  \BibitemOpen
  \bibfield  {author} {\bibinfo {author} {\bibfnamefont {A.}~\bibnamefont {L\"auchli}}\ and\ \bibinfo {author} {\bibfnamefont {D.}~\bibnamefont {Poilblanc}},\ }\bibfield  {title} {\bibinfo {title} {Spin-charge separation in two-dimensional frustrated quantum magnets},\ }\href {https://doi.org/10.1103/PhysRevLett.92.236404} {\bibfield  {journal} {\bibinfo  {journal} {Phys. Rev. Lett.}\ }\textbf {\bibinfo {volume} {92}},\ \bibinfo {pages} {236404} (\bibinfo {year} {2004})}\BibitemShut {NoStop}%
\bibitem [{\citenamefont {Miao}\ \emph {et~al.}(2025)\citenamefont {Miao}, \citenamefont {Yue}, \citenamefont {Zhang}, \citenamefont {Chen},\ and\ \citenamefont {Gu}}]{miao2025}%
  \BibitemOpen
  \bibfield  {author} {\bibinfo {author} {\bibfnamefont {J.-J.}\ \bibnamefont {Miao}}, \bibinfo {author} {\bibfnamefont {Z.-Y.}\ \bibnamefont {Yue}}, \bibinfo {author} {\bibfnamefont {H.}~\bibnamefont {Zhang}}, \bibinfo {author} {\bibfnamefont {W.-Q.}\ \bibnamefont {Chen}},\ and\ \bibinfo {author} {\bibfnamefont {Z.-C.}\ \bibnamefont {Gu}},\ }\bibfield  {title} {\bibinfo {title} {Spin-charge separation and unconventional superconductivity in the $t\text{\ensuremath{-}}{J}$ model on the honeycomb lattice},\ }\href {https://doi.org/10.1103/PhysRevB.111.174518} {\bibfield  {journal} {\bibinfo  {journal} {Phys. Rev. B}\ }\textbf {\bibinfo {volume} {111}},\ \bibinfo {pages} {174518} (\bibinfo {year} {2025})}\BibitemShut {NoStop}%
\bibitem [{\citenamefont {Yang}\ \emph {et~al.}(2024)\citenamefont {Yang}, \citenamefont {Devereaux},\ and\ \citenamefont {Jiang}}]{my_3band}%
  \BibitemOpen
  \bibfield  {author} {\bibinfo {author} {\bibfnamefont {L.}~\bibnamefont {Yang}}, \bibinfo {author} {\bibfnamefont {T.~P.}\ \bibnamefont {Devereaux}},\ and\ \bibinfo {author} {\bibfnamefont {H.-C.}\ \bibnamefont {Jiang}},\ }\bibfield  {title} {\bibinfo {title} {Recovery of a luther-emery phase in the three-band hubbard ladder with longer-range hopping},\ }\href {https://doi.org/10.1103/PhysRevB.110.014511} {\bibfield  {journal} {\bibinfo  {journal} {Phys. Rev. B}\ }\textbf {\bibinfo {volume} {110}},\ \bibinfo {pages} {014511} (\bibinfo {year} {2024})}\BibitemShut {NoStop}%
\bibitem [{\citenamefont {Gros}\ \emph {et~al.}(1987)\citenamefont {Gros}, \citenamefont {Joynt},\ and\ \citenamefont {Rice}}]{mapping_hubbard_tJ}%
  \BibitemOpen
  \bibfield  {author} {\bibinfo {author} {\bibfnamefont {C.}~\bibnamefont {Gros}}, \bibinfo {author} {\bibfnamefont {R.}~\bibnamefont {Joynt}},\ and\ \bibinfo {author} {\bibfnamefont {T.~M.}\ \bibnamefont {Rice}},\ }\bibfield  {title} {\bibinfo {title} {Antiferromagnetic correlations in almost-localized fermi liquids},\ }\href {https://doi.org/10.1103/PhysRevB.36.381} {\bibfield  {journal} {\bibinfo  {journal} {Phys. Rev. B}\ }\textbf {\bibinfo {volume} {36}},\ \bibinfo {pages} {381} (\bibinfo {year} {1987})}\BibitemShut {NoStop}%
\bibitem [{\citenamefont {White}(1992)}]{White1992}%
  \BibitemOpen
  \bibfield  {author} {\bibinfo {author} {\bibfnamefont {S.~R.}\ \bibnamefont {White}},\ }\bibfield  {title} {\bibinfo {title} {Density matrix formulation for quantum renormalization groups},\ }\href {https://doi.org/10.1103/PhysRevLett.69.2863} {\bibfield  {journal} {\bibinfo  {journal} {Phys. Rev. Lett.}\ }\textbf {\bibinfo {volume} {69}},\ \bibinfo {pages} {2863} (\bibinfo {year} {1992})}\BibitemShut {NoStop}%
\bibitem [{\citenamefont {White}(1993)}]{white1993}%
  \BibitemOpen
  \bibfield  {author} {\bibinfo {author} {\bibfnamefont {S.~R.}\ \bibnamefont {White}},\ }\bibfield  {title} {\bibinfo {title} {Density-matrix algorithms for quantum renormalization groups},\ }\href {https://doi.org/10.1103/PhysRevB.48.10345} {\bibfield  {journal} {\bibinfo  {journal} {Phys. Rev. B}\ }\textbf {\bibinfo {volume} {48}},\ \bibinfo {pages} {10345} (\bibinfo {year} {1993})}\BibitemShut {NoStop}%
\bibitem [{\citenamefont {White}\ and\ \citenamefont {Feiguin}(2004)}]{white2004}%
  \BibitemOpen
  \bibfield  {author} {\bibinfo {author} {\bibfnamefont {S.~R.}\ \bibnamefont {White}}\ and\ \bibinfo {author} {\bibfnamefont {A.~E.}\ \bibnamefont {Feiguin}},\ }\bibfield  {title} {\bibinfo {title} {Real-time evolution using the density matrix renormalization group},\ }\href {https://doi.org/10.1103/PhysRevLett.93.076401} {\bibfield  {journal} {\bibinfo  {journal} {Phys. Rev. Lett.}\ }\textbf {\bibinfo {volume} {93}},\ \bibinfo {pages} {076401} (\bibinfo {year} {2004})}\BibitemShut {NoStop}%
\bibitem [{\citenamefont {Daley}\ \emph {et~al.}(2004)\citenamefont {Daley}, \citenamefont {Kollath}, \citenamefont {Schollw\"ock},\ and\ \citenamefont {Vidal}}]{daley2004}%
  \BibitemOpen
  \bibfield  {author} {\bibinfo {author} {\bibfnamefont {A.~J.}\ \bibnamefont {Daley}}, \bibinfo {author} {\bibfnamefont {C.}~\bibnamefont {Kollath}}, \bibinfo {author} {\bibfnamefont {U.}~\bibnamefont {Schollw\"ock}},\ and\ \bibinfo {author} {\bibfnamefont {G.}~\bibnamefont {Vidal}},\ }\bibfield  {title} {\bibinfo {title} {Time-dependent density-matrix renormalization-group using adaptive effective hilbert spaces},\ }\href {https://doi.org/10.1088/1742-5468/2004/04/p04005} {\bibfield  {journal} {\bibinfo  {journal} {Journal of Statistical Mechanics: Theory and Experiment}\ }\textbf {\bibinfo {volume} {2004}},\ \bibinfo {pages} {P04005} (\bibinfo {year} {2004})}\BibitemShut {NoStop}%
\bibitem [{\citenamefont {Feiguin}(2011)}]{vietri}%
  \BibitemOpen
  \bibfield  {author} {\bibinfo {author} {\bibfnamefont {A.~E.}\ \bibnamefont {Feiguin}},\ }\bibfield  {title} {\bibinfo {title} {The density matrix renormalization group method and its time-dependent variants},\ }in\ \href@noop {} {\emph {\bibinfo {booktitle} {XV Training Course in the Physics of Strongly Correlated Systems}}},\ Vol.\ \bibinfo {volume} {1419}\ (\bibinfo  {publisher} {AIP Proceedings},\ \bibinfo {year} {2011})\ p.~\bibinfo {pages} {5}\BibitemShut {NoStop}%
\bibitem [{\citenamefont {Paeckel}\ \emph {et~al.}(2019)\citenamefont {Paeckel}, \citenamefont {Köhler}, \citenamefont {Swoboda}, \citenamefont {Manmana}, \citenamefont {Schollwöck},\ and\ \citenamefont {Hubig}}]{Paeckel2019}%
  \BibitemOpen
  \bibfield  {author} {\bibinfo {author} {\bibfnamefont {S.}~\bibnamefont {Paeckel}}, \bibinfo {author} {\bibfnamefont {T.}~\bibnamefont {Köhler}}, \bibinfo {author} {\bibfnamefont {A.}~\bibnamefont {Swoboda}}, \bibinfo {author} {\bibfnamefont {S.~R.}\ \bibnamefont {Manmana}}, \bibinfo {author} {\bibfnamefont {U.}~\bibnamefont {Schollwöck}},\ and\ \bibinfo {author} {\bibfnamefont {C.}~\bibnamefont {Hubig}},\ }\bibfield  {title} {\bibinfo {title} {Time-evolution methods for matrix-product states},\ }\href {https://doi.org/https://doi.org/10.1016/j.aop.2019.167998} {\bibfield  {journal} {\bibinfo  {journal} {Annals of Physics}\ }\textbf {\bibinfo {volume} {411}},\ \bibinfo {pages} {167998} (\bibinfo {year} {2019})}\BibitemShut {NoStop}%
\bibitem [{\citenamefont {Noack}\ \emph {et~al.}(1996)\citenamefont {Noack}, \citenamefont {White},\ and\ \citenamefont {Scalapino}}]{NOACK1996281}%
  \BibitemOpen
  \bibfield  {author} {\bibinfo {author} {\bibfnamefont {R.}~\bibnamefont {Noack}}, \bibinfo {author} {\bibfnamefont {S.}~\bibnamefont {White}},\ and\ \bibinfo {author} {\bibfnamefont {D.}~\bibnamefont {Scalapino}},\ }\bibfield  {title} {\bibinfo {title} {The ground state of the two-leg hubbard ladder a density-matrix renormalization group study},\ }\href {https://doi.org/https://doi.org/10.1016/S0921-4534(96)00515-1} {\bibfield  {journal} {\bibinfo  {journal} {Physica C: Superconductivity}\ }\textbf {\bibinfo {volume} {270}},\ \bibinfo {pages} {281} (\bibinfo {year} {1996})}\BibitemShut {NoStop}%
\bibitem [{\citenamefont {Yang}\ and\ \citenamefont {Feiguin}(2019)}]{Adrian_hubbard_ladder}%
  \BibitemOpen
  \bibfield  {author} {\bibinfo {author} {\bibfnamefont {C.}~\bibnamefont {Yang}}\ and\ \bibinfo {author} {\bibfnamefont {A.~E.}\ \bibnamefont {Feiguin}},\ }\bibfield  {title} {\bibinfo {title} {Spectral function of mott-insulating hubbard ladders: From fractionalized excitations to coherent quasiparticles},\ }\href {https://doi.org/10.1103/PhysRevB.99.235117} {\bibfield  {journal} {\bibinfo  {journal} {Phys. Rev. B}\ }\textbf {\bibinfo {volume} {99}},\ \bibinfo {pages} {235117} (\bibinfo {year} {2019})}\BibitemShut {NoStop}%
\bibitem [{\citenamefont {Ogata}\ and\ \citenamefont {Shiba}(1990)}]{ogata-shiba}%
  \BibitemOpen
  \bibfield  {author} {\bibinfo {author} {\bibfnamefont {M.}~\bibnamefont {Ogata}}\ and\ \bibinfo {author} {\bibfnamefont {H.}~\bibnamefont {Shiba}},\ }\bibfield  {title} {\bibinfo {title} {Bethe-ansatz wave function, momentum distribution, and spin correlation in the one-dimensional strongly correlated hubbard model},\ }\href {https://doi.org/10.1103/PhysRevB.41.2326} {\bibfield  {journal} {\bibinfo  {journal} {Phys. Rev. B}\ }\textbf {\bibinfo {volume} {41}},\ \bibinfo {pages} {2326} (\bibinfo {year} {1990})}\BibitemShut {NoStop}%
\bibitem [{\citenamefont {Yang}\ \emph {et~al.}(2022)\citenamefont {Yang}, \citenamefont {Hamad}, \citenamefont {Manuel},\ and\ \citenamefont {Feiguin}}]{my_tJ_chain}%
  \BibitemOpen
  \bibfield  {author} {\bibinfo {author} {\bibfnamefont {L.}~\bibnamefont {Yang}}, \bibinfo {author} {\bibfnamefont {I.}~\bibnamefont {Hamad}}, \bibinfo {author} {\bibfnamefont {L.~O.}\ \bibnamefont {Manuel}},\ and\ \bibinfo {author} {\bibfnamefont {A.~E.}\ \bibnamefont {Feiguin}},\ }\bibfield  {title} {\bibinfo {title} {Fate of pairing and spin-charge separation in the presence of long-range antiferromagnetism},\ }\href {https://doi.org/10.1103/PhysRevB.105.195104} {\bibfield  {journal} {\bibinfo  {journal} {Phys. Rev. B}\ }\textbf {\bibinfo {volume} {105}},\ \bibinfo {pages} {195104} (\bibinfo {year} {2022})}\BibitemShut {NoStop}%
\bibitem [{\citenamefont {Nocera}\ \emph {et~al.}(2018)\citenamefont {Nocera}, \citenamefont {Essler},\ and\ \citenamefont {Feiguin}}]{Adrian_finite_temperature}%
  \BibitemOpen
  \bibfield  {author} {\bibinfo {author} {\bibfnamefont {A.}~\bibnamefont {Nocera}}, \bibinfo {author} {\bibfnamefont {F.~H.~L.}\ \bibnamefont {Essler}},\ and\ \bibinfo {author} {\bibfnamefont {A.~E.}\ \bibnamefont {Feiguin}},\ }\bibfield  {title} {\bibinfo {title} {Finite-temperature dynamics of the mott insulating hubbard chain},\ }\href {https://doi.org/10.1103/PhysRevB.97.045146} {\bibfield  {journal} {\bibinfo  {journal} {Phys. Rev. B}\ }\textbf {\bibinfo {volume} {97}},\ \bibinfo {pages} {045146} (\bibinfo {year} {2018})}\BibitemShut {NoStop}%
\bibitem [{\citenamefont {Yang}\ \emph {et~al.}(2026)\citenamefont {Yang}, \citenamefont {Feiguin}, \citenamefont {Devereaux},\ and\ \citenamefont {Dagotto}}]{yang2026effects}%
  \BibitemOpen
  \bibfield  {author} {\bibinfo {author} {\bibfnamefont {L.}~\bibnamefont {Yang}}, \bibinfo {author} {\bibfnamefont {A.~E.}\ \bibnamefont {Feiguin}}, \bibinfo {author} {\bibfnamefont {T.~P.}\ \bibnamefont {Devereaux}},\ and\ \bibinfo {author} {\bibfnamefont {E.}~\bibnamefont {Dagotto}},\ }\bibfield  {title} {\bibinfo {title} {Effects of the next-nearest-neighbor hopping on the low-dimensional hubbard model: ferromagnetism, antiferromagnetism, and superconductivity},\ }\href@noop {} {\bibfield  {journal} {\bibinfo  {journal} {Journal of Physics: Condensed Matter}\ }\textbf {\bibinfo {volume} {38}},\ \bibinfo {pages} {023002} (\bibinfo {year} {2026})}\BibitemShut {NoStop}%
\bibitem [{\citenamefont {Hellberg}\ and\ \citenamefont {Mele}(1993)}]{others}%
  \BibitemOpen
  \bibfield  {author} {\bibinfo {author} {\bibfnamefont {C.~S.}\ \bibnamefont {Hellberg}}\ and\ \bibinfo {author} {\bibfnamefont {E.~J.}\ \bibnamefont {Mele}},\ }\bibfield  {title} {\bibinfo {title} {Luttinger-liquid instability in the one-dimensional t-j model},\ }\href {https://doi.org/10.1103/PhysRevB.48.646} {\bibfield  {journal} {\bibinfo  {journal} {Phys. Rev. B}\ }\textbf {\bibinfo {volume} {48}},\ \bibinfo {pages} {646} (\bibinfo {year} {1993})},\ \bibinfo {note} {where a transition from TLL to Luther-Emery was also observed for the $t-{J}$ model by numerically projecting the true ground state from a Luttinger liquid wave function, varying $J/t$ and electronic density, and focusing on the crossover from linear to quadratic behavior at small wave vectors in the spin-spin correlation function $S({\bf k})$.}\BibitemShut {Stop}%
\bibitem [{\citenamefont {Laurell}\ \emph {et~al.}(2024)\citenamefont {Laurell}, \citenamefont {Herbrych}, \citenamefont {Alvarez},\ and\ \citenamefont {Dagotto}}]{pontus}%
  \BibitemOpen
  \bibfield  {author} {\bibinfo {author} {\bibfnamefont {P.}~\bibnamefont {Laurell}}, \bibinfo {author} {\bibfnamefont {J.}~\bibnamefont {Herbrych}}, \bibinfo {author} {\bibfnamefont {G.}~\bibnamefont {Alvarez}},\ and\ \bibinfo {author} {\bibfnamefont {E.}~\bibnamefont {Dagotto}},\ }\bibfield  {title} {\bibinfo {title} {Luther-emery liquid and dominant singlet superconductivity in the hole-doped haldane spin-1 chain},\ }\href {https://doi.org/10.1103/PhysRevB.110.064515} {\bibfield  {journal} {\bibinfo  {journal} {Phys. Rev. B}\ }\textbf {\bibinfo {volume} {110}},\ \bibinfo {pages} {064515} (\bibinfo {year} {2024})}\BibitemShut {NoStop}%
\bibitem [{\citenamefont {Patel}\ \emph {et~al.}(2017)\citenamefont {Patel}, \citenamefont {Nocera}, \citenamefont {Alvarez}, \citenamefont {Moreo},\ and\ \citenamefont {Dagotto}}]{patel1}%
  \BibitemOpen
  \bibfield  {author} {\bibinfo {author} {\bibfnamefont {N.~D.}\ \bibnamefont {Patel}}, \bibinfo {author} {\bibfnamefont {A.}~\bibnamefont {Nocera}}, \bibinfo {author} {\bibfnamefont {G.}~\bibnamefont {Alvarez}}, \bibinfo {author} {\bibfnamefont {A.}~\bibnamefont {Moreo}},\ and\ \bibinfo {author} {\bibfnamefont {E.}~\bibnamefont {Dagotto}},\ }\bibfield  {title} {\bibinfo {title} {Pairing tendencies in a two-orbital hubbard model in one dimension},\ }\href {https://doi.org/10.1103/PhysRevB.96.024520} {\bibfield  {journal} {\bibinfo  {journal} {Phys. Rev. B}\ }\textbf {\bibinfo {volume} {96}},\ \bibinfo {pages} {024520} (\bibinfo {year} {2017})}\BibitemShut {NoStop}%
\bibitem [{\citenamefont {Patel}\ \emph {et~al.}(2020)\citenamefont {Patel}, \citenamefont {Kaushal}, \citenamefont {Nocera}, \citenamefont {Alvarez},\ and\ \citenamefont {Dagotto}}]{patel2}%
  \BibitemOpen
  \bibfield  {author} {\bibinfo {author} {\bibfnamefont {N.~D.}\ \bibnamefont {Patel}}, \bibinfo {author} {\bibfnamefont {N.}~\bibnamefont {Kaushal}}, \bibinfo {author} {\bibfnamefont {A.}~\bibnamefont {Nocera}}, \bibinfo {author} {\bibfnamefont {G.}~\bibnamefont {Alvarez}},\ and\ \bibinfo {author} {\bibfnamefont {E.}~\bibnamefont {Dagotto}},\ }\bibfield  {title} {\bibinfo {title} {Emergence of superconductivity in doped multiorbital hubbard chains},\ }\href@noop {} {\bibfield  {journal} {\bibinfo  {journal} {npj Quantum Materials}\ }\textbf {\bibinfo {volume} {5}},\ \bibinfo {pages} {27} (\bibinfo {year} {2020})}\BibitemShut {NoStop}%
\end{thebibliography}%
\end{document}